\documentclass[11pt]{article}
\pdfoutput=1

\usepackage[colorlinks,urlcolor=black,linkcolor = blue,citecolor = black,]{hyperref}
\usepackage[skip=0.3em,indent]{parskip}
\usepackage{subfigure}
\usepackage{latexsym}
\usepackage{epsfig}
\usepackage[mathscr]{eucal}
\usepackage{amsfonts}
\usepackage{amscd}
\usepackage{cite}
\usepackage{array}
\usepackage{amssymb}
\usepackage{colordvi}
\usepackage[centertags]{amsmath}
\usepackage{enumerate}
\usepackage{graphicx}
\usepackage{booktabs}
\usepackage{theorem}
\usepackage[footnotesize]{caption}
\usepackage{soul}
\usepackage{mcite}
\usepackage{slashed}
\usepackage{braket}
\usepackage{units}

\usepackage[dvipsnames]{xcolor}
\usepackage{ulem}
\usepackage{bbm}
\usepackage[utf8]{inputenc}
\usepackage{fancyvrb}
\usepackage{framed}
\usepackage{xspace}
\usepackage[textsize=footnotesize]{todonotes}

\newcommand{\newc}{\newcommand}

\allowdisplaybreaks
\usepackage[printonlyused]{acronym}
\newcommand{\caO}{c_{\alpha_1}}
\newcommand{\saO}{s_{\alpha_1}}
\newcommand{\caT}{c_{\alpha_2}}
\newcommand{\saT}{s_{\alpha_2}}
\newcommand{\caTh}{c_{\alpha_3}}
\newcommand{\saTh}{s_{\alpha_3}}
\newcommand{\caF}{c_{\alpha_4}}
\newcommand{\saF}{s_{\alpha_4}}
\newcommand{\caFi}{c_{\alpha_5}}
\newcommand{\saFi}{s_{\alpha_5}}
\newcommand{\caS}{c_{\alpha_6}}
\newcommand{\saS}{s_{\alpha_6}}
\def\<#1>{\mathinner{\langle#1\rangle}}

\newc{\bea}{\begin{eqnarray}}
\newc{\eea}{\end{eqnarray}}
\newc{\ol}{\overline}
\newc{\wt}{\widetilde}
\newc{\bs}{\boldsymbol}
\newc{\m}{\mathcal}
\newc{\la}{\langle}
\newc{\ra}{\rangle}

\newcommand{\nn}{\nonumber}
\newcommand{\bpmatrix}{\begin{pmatrix}}
\newcommand{\epmatrix}{\end{pmatrix}}
\newcommand{\ba}{\begin{array}}
\newcommand{\ea}{\end{array}}

\renewcommand{\ol}{\text{1l}}
\renewcommand{\Re}{\text{Re}\!}
\renewcommand{\Im}{\text{Im}\!}

\renewcommand{\Re}{\operatorname{Re}}
\renewcommand{\Im}{\operatorname{Im}}

\newcommand{\ie}{\textit{i.e.}\ }
\newcommand{\eg}{\textit{e.g.}\ }
\newcommand{\bc}{\begin{center}}
\newcommand{\ec}{\end{center}}

\newcommand{\gev}{~\text{GeV}}

\newcommand{\fb}{{~\text{fb}}}

\newcommand{\beq}{\begin{eqnarray}}
\newcommand{\eeq}{\end{eqnarray}}

\newenvironment{kasten*}[1]
{
\hspace{0.05\linewidth}
\begin{minipage}{0.95\linewidth}
\setlength{\fboxsep}{10pt}
\definecolor{shadecolor}{gray}{0.9}
\definecolor{framecolor}{gray}{0}

\MakeFramed {\FrameRestore}
\subsection*{#1}
}
{
\endMakeFramed
\end{minipage}
\vspace{1em}
}

\usepackage{cleveref}
\crefname{chapter}{Chapter}{Chapter}
\crefname{section}{Sec.}{Secs.}
\crefname{table}{Tab.}{Tabs.}
\crefname{figure}{Fig.}{Figs.}
\crefname{equation}{Eq.}{Eqs.}
\crefname{appendix}{Appendix\ }{Appendix\ }

\allowdisplaybreaks

\begin{document}
\title{
  \textbf{The CN2HDM \\[4mm]}}

\date{}

\author{
Margarete M\"{u}hlleitner$^{1\,}$\footnote{\texttt{margarete.muehlleitner@kit.edu}} ,
Jonas M\"{u}ller$^{1\,}$\footnote{\texttt{jonas.mueller@kit.edu}},
Sophie L. Williamson$^{1\,}$\footnote{\texttt{sophie.williamson@kit.edu}} ,
Jonas Wittbrodt $^{2}$\footnote{\texttt{jonas.wittbrodt@desy.de}}
\\[9mm]
{\small\it
$^1$Institute for Theoretical Physics, Karlsruhe Institute of Technology,} \\
{\small\it 76128 Karlsruhe, Germany}\\
{\small\it
$^2$Department of Astronomy and Theoretical Physics, Lund University, S\"olvegatan 14A,} \\
{\small\it 223 62 Lund, Sweden}
}
%
%
\maketitle

\begin{flushright}
  \vspace{-8cm}
  KA-TP-20-2021\\
  LU TP 21-39
  \vspace{7cm}
\end{flushright}

\begin{abstract}
  We present the CP-violating Next-to-2-Higgs-Doublet Model (CN2HDM)
  which is based on the extension of the CP-violating
  2-Higgs-Doublet-Model (C2HDM) by a complex singlet field that
  obeys a discrete $\mathbb{Z}_2$ symmetry. The model thus features not
  only CP violation required for successful electroweak baryogenesis but also a Dark
  Matter (DM)
  candidate. The model has an extended Higgs sector with four
  CP-mixing visible neutral Higgs bosons, a DM candidate and a pair of
  oppositely charged Higgs bosons. The possibility of singlet and
  CP-odd admixtures to the observed Higgs boson in addition to the
  large number of visible scalar particles leads to an interesting
  Higgs phenomenology. We find that the model can easily provide 100\%
  of the DM relic density and investigate interesting LHC and DM
  observables within the model. We provide all the tools necessary to
  study the CN2HDM in detail and point out future research directions
  for this interesting benchmark model that can address some of the most pressing
  open questions of the Standard Model.
\end{abstract}
\thispagestyle{empty}
\vfill
\newpage
\setcounter{page}{1}

\maketitle

\section{Introduction}

The discovery of a Higgs boson by the ATLAS
\cite{Aad:2012tfa} and CMS \cite{Chatrchyan:2012xdj} collaborations
that behaves very similar to the expectations within the Standard
Model (SM) has undoubtedly been the founding achievement of the LHC
runs so far. Contrarily, searches for physics beyond the Standard
Model (BSM) have come up empty so far, but remain promising
avenues to explore. BSM physics is motivated
by a number of questions unanswered by the SM, including those on how to
explain Dark Matter (DM) and Dark Energy that make up $95\%$~\cite{Planck:2015fie, ParticleDataGroup:2016lqr, Planck:2015mrs} of
the physical composition of our universe or how the observed
baryon-antibaryon asymmetry~\cite{Bennett:2012zja} can be established. A
dynamical mechanism for the generation of the asymmetry is
provided by electroweak baryogenesis (EWBG)~\cite{Kuzmin:1985mm,Cohen:1990it,Cohen:1993nk,Quiros:1994dr,Rubakov:1996vz,Funakubo:1996dw,Trodden:1998ym,Bernreuther:2002uj,Morrissey:2012db}, which requires the three
Sakharov conditions to be fulfilled~\cite{Sakharov:1967dj}. Among
these is the necessity of sufficiently large CP violation which requires
additional sources of CP violation beyond the SM.

There is no current
precondition to suggest the form that DM should take,
whether it be made up of axions, weakly interactive massive particles
(WIMPs), or primordial black holes, to name only a few. It is not hard
to construct extensions of the SM Higgs sector that can
account for the entire DM relic density of the universe through
WIMPs. This can be achieved \eg by adding a neutral real or
complex singlet field~\cite{Silveira:1985rk, McDonald:1993ex, Burgess:2000yq}.
Another way is the addition of a second $SU(2)_L$ doublet
of hypercharge $Y=\pm 1$, which keeps the electroweak precision
parameter $\rho=1$ as in the SM \cite{Lee:1973iz,
  ParticleDataGroup:2018ovx} and --- with its two complex Higgs
doublets --- reflects the Higgs sector that is the basis for
supersymmetry. This extension of the SM Higgs sector gives rise to
variations of the 2-Higgs Doublet Model (2HDM) \cite{Lee:1973iz,Branco:2011iw},
classified by the manner in which the Higgs doublets couple to the fermions.
Dark Matter can arise from 2HDMs in various ways, for example in the inert
2HDM~\cite{Deshpande:1977rw, Ma:2006km, Barbieri:2006dq,
  LopezHonorez:2006gr, Lundstrom:2008ai, Dolle:2009fn, LopezHonorez:2010tb,
  LopezHonorez:2010eeh, Goudelis:2013uca,
  Bonilla:2014xba, Queiroz:2015utg} where one of the two Higgs
doublets is inert, \ie it does not obtain a vacuum expectation
value and hence provides a scalar DM candidate.

CP violation can be induced in the Higgs sector of the 2HDM, without
the addition of new fermions, by constructing a 2HDM potential with a
softly broken $\mathbb{Z}_2$ symmetry with two complex parameters,
$\lambda_5$ and $m_{12}^2$ \cite{Ginzburg:2002wt, Khater:2003wq,
  ElKaffas:2007rq, Grzadkowski:2009iz, Arhrib:2010ju, Barroso:2012wz,
  Inoue:2014nva, Cheung:2014oaa, Fontes:2014xva, Fontes:2015mea,
  Chen:2015gaa, Fontes:2017zfn, Boto:2020wyf, Fontes:2021iue}.
Besides CP violation, another salient element for successful EWBG is the
interplay between the Higgs mass spectrum and the self-interactions of
the Higgs bosons participating in the electroweak phase transition
(EWPT). Therefore, a major goal of the upcoming high-luminosity Large
Hadron Collider (LHC) is the measurement of the Higgs self-couplings,
which is a non-trivial task and requires large di-Higgs cross sections
stemming from resonant production of heavy Higgs bosons. The 2HDM
becomes increasingly constrained with respect to these collider
processes. Additionally, in its type II version it only allows for a heavy Higgs
spectrum. This is different in the Next-to-2HDM (N2HDM) \cite{Chen:2013jvg, Drozd:2014yla, Muhlleitner:2016mzt}.

In the N2HDM, an additional real singlet field is added to the two
Higgs doublets. The N2HDM is
invariant under two $\mathbb{Z}_2$ symmetries and, depending on the
way in which they are spontaneously broken, the model yields different
dark phases with different numbers of DM candidates (or none
at all)~\cite{Engeln:2020fld}. Furthermore, the Higgs mass eigenstates
get some admixture from the singlet field so that light Higgs states
are still allowed in the spectrum that have escaped discovery so
far. This implies an interesting LHC
phenomenology~\cite{Muhlleitner:2016mzt,Ferreira:2019iqb,Engeln:2020fld,Azevedo:2021ylf} including the
possibility of resonant Higgs production with subsequent decays into
Higgs pair final states and even Higgs-to-Higgs cascade decays. The
extended Higgs sector can also allow for a strong
first order EWPT~\cite{Basler:2019iuu,Basler:2018cwe,Basler:2020nrq}
that is required for EWBG~\cite{Trodden:1998ym,Morrissey:2012db}.  
Thus, the N2HDM can be built up such that it can address some of
the most pressing open questions of the SM and additionally provide an
interesting landscape for LHC phenomenology. However, while the N2HDM
can accommodate DM and realise a first order EWPT, it is defined to be
CP-conserving and thus EWBG cannot be realised.

In this work, we present an extension of the N2HDM that addresses this
limitation: the CP-violating N2HDM (CN2HDM). It not only incorporates
CP violation into its Higgs potential but also contains a DM candidate
on top of the particle content of the N2HDM. To achieve this, the
model adds a complex singlet field --- instead of the real singlet
field of the N2HDM --- with the $\mathbb{Z}_2$ spontaneously broken
for the real part, but unbroken for the imaginary
component which becomes the DM candidate. At the same time, the real CP-even
parts of the doublet and the real singlet field have non-vanishing vacuum
expectation values.
This construction gives rise to four neutral visible Higgs bosons, a
pair of visible oppositely charged Higgs bosons and one DM
candidate, allowing for an extremely rich phenomenology. We impose on our
model both collider and DM constraints, as well as the constraints stemming from
the measurements of the electric dipole moments that are relevant for
CP violation. We subsequently investigate --- with viable parameter
points fulfilling all the imposed constraints --- the CN2HDM DM observables and
its phenomenology at the LHC. The aim of our paper is the
introduction of an interesting benchmark model that features an
extended Higgs sector with singlet admixture that is able to address
two pressing unsolved questions of contemporary particle physics, the
DM candidate and the matter-antimatter asymmetry puzzles. Being a
non-supersymmetric model it allows for more freedom in its parameter
space and comes along with an interesting LHC phenomenology that
can be tested at present and future LHC runs as well as future
$e^+e^-$ colliders.



The outline of our paper is as follows. In \cref{sec:model} we
introduce the CN2HDM with its field content and set up our notation.
In the subsequent numerical analysis in \cref{sec:numericalanalysis}, we
present the constraints applied and the details of the parameter
scan performed. For the latter, we implemented our model in {\tt ScannerS}~\cite{Coimbra:2013qq,Ferreira:2014dya,Muhlleitner:2016mzt,Muhlleitner:2020wwk}
and furthermore provide a new Fortran code, {\tt CN2HDM\_HDECAY}, for the
computation of the decay widths and branching ratios including
state-of-the-art higher-order corrections.
We then discuss the interplay of DM constraints and DM
observables in our model in \cref{sec:DMconstraints} before we
move on to the investigation of the CN2HDM
Higgs boson phenomenology at the LHC both in the type I and type II
versions of the model in \cref{sec:higgspheno}. We summarise the
findings of our analysis in the
conclusion and give an outlook for future directions. In the Appendix,
we present auxiliary material, including information on the mixing
matrix and physical parameter relations.



\section{Model Introduction}\label{sec:model}
The starting point of our CN2HDM model is the N2HDM
\cite{Chen:2013jvg, Drozd:2014yla, Muhlleitner:2016mzt} that we adapt
such that it includes both a DM candidate and CP
violation. The CN2HDM Higgs sector hence consists of the SM extended
by a complex $SU(2)_L$ doublet with hypercharge $Y=1$, and a complex
$U(1)_Y$ singlet with $Y=0$. The potential has two $\mathbb{Z}_2$
symmetries: $\Phi_{1,S} \to \Phi_{1,S}, \Phi_2 \to - \Phi_2$ is softly
broken by $m_{12}^2$, while $\Phi_{1,2}\to\Phi_{1,2}, \Phi_S \to
  -\Phi_S$ remains unbroken in the scalar potential. The potential is
given by
\beq
V &=& m_{11}^2 \Phi_1^\dagger \Phi_1 + m_{22}^2 \Phi_2^\dagger
\Phi_2 + \frac{\lambda_1}{2} \left( \Phi_1^\dagger
\Phi_1\right)^2 + \frac{\lambda_2}{2}
\left(\Phi_2^\dagger \Phi_2\right)^\dagger + \lambda_3
\Phi_1^\dagger \Phi_1 \Phi_2^\dagger \Phi_2 + \lambda_4
\Phi_1^\dagger \Phi_2 \Phi_2^\dagger \Phi_1 \nonumber \\
&&+ \frac{\lambda_6}{8} \vert S\vert^4 + \frac{\lambda_7}{2}
\Phi_1^\dagger \Phi_1 \vert S \vert^2 + \frac{\lambda_8}{2}
\Phi_2^\dagger \Phi_2 \vert S \vert^2 + \frac{1}{2} m_s^2 \vert S
\vert^2 \nonumber \\
&& + \left(\frac{\lambda_5}{2} \left( \Phi_1^\dagger \Phi_2 \right)^2 -
m_{12}^2 \Phi_1^\dagger \Phi_2 + c.c.\right) -
\frac{m_{\text{DM}}^2}{4} \left( S^2 + \left( S^\ast \right)^2 \right) \,.
\eeq
The parameters of the scalar potential are real except for the
complex $\lambda_5$ and $m_{12}^2$, which induce
CP violation in the model. The last term proportional to
$m_{\text{DM}}^2$ is new compared to the usual N2HDM potential
\cite{Muhlleitner:2016mzt}
and introduces a DM candidate.\footnote{Note, that a DM candidate can
  also be realised without introducing this additional term. This would, however, require, that either the singlet state $\Phi_S$ or one of the doublets does not acquire a vacuum expectation value (VEV)~\cite{Engeln:2020fld}. We do not pursue that direction here, since we want to retain the rich visible sector phenomenology with both CP and singlet mixing.}
The doublet $\Phi_i (i =1,2)$ and singlet $\Phi_S$ fields are further
decomposed into component fields as
\begin{align}
  \Phi_i = & \frac{1}{\sqrt{2}}  \begin{pmatrix}
    \zeta_i + i \eta_i \\ \rho_i + v_i + i \psi_i
  \end{pmatrix} \,, \nonumber \\
  \Phi_S = & \frac{1}{\sqrt{2}} \left( s + v_s + i a \right) \,,
\end{align}
where $\zeta_i$, $\eta_i$, $\rho_i$ and $\psi_i$ are the charged
CP-even, charged CP-odd, neutral CP-even and neutral CP-odd components
of the two doublet fields, respectively,
$s$ and $a$ the real and imaginary components of the singlet field,
respectively, and $v_I$ $(I = 1, 2, s)$ the VEVs of the
doublet and singlet fields.
We allow only the real singlet field to acquire a VEV\@. This breaks
the $\Phi_S\to -\Phi_S$ $\mathbb{Z}_2$ symmetry of the scalar
potential down to $\Phi_S\to\Phi_S^*$ --- or equivalently $a\to
  -a$. This remaining $\mathbb{Z}_2$ symmetry is what stabilises the DM
candidate field $a$.
Expanding around the minimum of the scalar potential given by the
VEVs, we obtain the tadpole conditions for electroweak symmetry breaking (EWSB),
\begin{subequations}
  \begin{align}
    \Im m_{12}^2 & = \Im \lambda_5 \frac{v_1 v_2}{2}  \label{Tad1}                                                                                                              \\
    m_{11}^2     & = -\frac{1}{2} \lambda_1 v_1^2  -\frac{1}{2} \left( \lambda_3 + \lambda_4 + \Re \lambda_5 \right) v_2^2 - \frac{\lambda_7}{4} v_s^2 + \Re m_{12}^2 \tan\beta \\
    m_{22}^2     & = - \frac{1}{2} \lambda_2 v_2^2 - \frac{1}{2} \left(\lambda_3+\lambda_4 + \Re \lambda_5 \right) v_1^2 - \frac{\lambda_8}{4} v_s^2   + \Re m_{12}^2 \tan\beta \\
    m_s^2        & = - \frac{\lambda_6}{4} v_s^2 - \frac{\lambda_7}{2} v_1^2 - \frac{\lambda_8}{2} v_2^2  + m_{\text{DM}}^2 \,,
  \end{align}
\end{subequations}
with
\begin{equation}
  \tan\beta = \frac{v_2}{v_1}\,,
\end{equation} where $v_1$ and $v_2$ relate to the SM-Higgs VEV $v\approx
  246$~GeV via $v = \sqrt{v_1^2 + v_2^2}$.

There are two CP-violating phases, $\phi(\lambda_5)$ and
$\phi(m_{12}^2)$, associated with the complex scalar potential
parameters, which we define as
\begin{align}
  \lambda_5 = \left| \lambda_5 \right| e^{i\phi(\lambda_5)} \qquad
  \mbox{and} \qquad m_{12}^2 = \left| m_{12}^2 \right|
  e^{i\phi(m_{12}^2)}\,.
\end{align}
As in the complex version of the
2HDM\cite{Fontes:2017zfn,Ginzburg:2002wt}, these two phases are not
independent, and the tadpole equation in \cref{Tad1} can be rewritten
as
\begin{align}
  2 \Re (m_{12}^2) \tan\left( \phi(m_{12}^2) \right) = v_1 v_2
  \Re(\lambda_5)\tan\left(\phi(\lambda_5)\right)\,.
\end{align}
We choose both vacuum expectation values $v_1$ and
  $v_2$ to be real. Together with the condition $\phi(\lambda_5) \ne 2
  \phi(m_{12}^2)$ \cite{Ginzburg:2002wt} this ensures that the two phases cannot be
  removed simultaneously, which would bring us back to the
  CP-conserving limit of the model. 
The charged sector is diagonalised via
\begin{align}
  \begin{pmatrix}
    G^\pm \\ H^\pm
  \end{pmatrix} = & \begin{pmatrix}
    \cos\beta & \sin\beta \\ -\sin \beta & \cos\beta
  \end{pmatrix} \begin{pmatrix}
    \frac{1}{\sqrt{2}} \left(\zeta_1 \pm i \eta_1 \right) \\
    \frac{1}{\sqrt{2}} \left(\zeta_2 \pm i \eta_2\right)
  \end{pmatrix} \,,
\end{align}
which yields a charged massless Goldstone, $G^\pm$, like in the SM,
and a charged Higgs, $H^\pm$, with the mass
\begin{align}
  m_{H^\pm}^2 = - \frac{v^2}{2} \left(\lambda_4 + \Re \lambda_5 \right)
  + \frac{\Re m_{12}^2}{\sin\beta \cos\beta} \,.
\end{align}
The CP-odd component $a$ of the singlet $\Phi_S$ does not mix with the
other states, and yields a DM candidate with the mass $m_{\text{DM}}$.
The two neutral CP-odd components $\psi_{1,2}$ of the two doublets
$\Phi_{1,2}$ are first rotated to the neutral massless Goldstone boson
$G^0$ and a CP-odd field $\rho_3$,
\begin{align}
  \begin{pmatrix}
    G^0 \\
    \rho_3
  \end{pmatrix}
  = & \begin{pmatrix}
    \cos\beta & \sin\beta \\ -\sin \beta & \cos\beta
  \end{pmatrix}
  \begin{pmatrix}
    \psi_1 \\ \psi_2
  \end{pmatrix} \,.
\end{align}
Subsequently, the component fields $n_i$ = $\rho_1, \rho_2, s, \rho_3$
are rotated to the Higgs mass
eigenstates via 
\begin{align}
  \begin{pmatrix}
    H_a \\ H_b \\ H_c \\ H_d
  \end{pmatrix} = R \begin{pmatrix}
    \rho_1 \\ \rho_2 \\ s \\ \rho_3
  \end{pmatrix} \,,
\end{align}
where $R$ denotes the $4\times4$ mixing matrix that is parametrised by
six mixing angles $\alpha_{1,\ldots,6}$. Without loss of generality
they can be chosen to lie in the range
\begin{equation}
  -\frac{\pi}{2} \le \alpha_{1,\ldots,6}  \le \frac{\pi}{2} \;.
\end{equation}
The explicit form of $R$ is given
in \cref{Appendix:MixingMatrix}. The four neutral Higgs particles
have no definite CP quantum numbers and their masses are obtained
from the mass matrix
\begin{align}
  \left(\mathcal{M}_N^2\right)_{ij} = \biggl< \frac{\partial^2 V}{\partial n_i \partial n_j} \biggr> \,,
\end{align} with
\begin{align}
  R \mathcal{M}_N^2 R^T = & \mathrm{diag}\left(m_{H_a}^2,m_{H_b}^2,m_{H_c}^2 ,m_{H_d}^2 \right)\,,
\end{align}
where $H_{a,b,c,d}$ denote the mass eigenstates.

In order to avoid tree-level flavour-changing neutral currents (FCNC)
the softly broken $\mathbb{Z}_2$ symmetry acting on the doublet fields
is extended to the Yukawa sector. Just
like in the 2HDM this implies four different types of doublet
couplings to the fermions, which are listed in \cref{tab:yuycoup}.
\begin{table}
  \begin{center}
    \begin{tabular}{rccc|ccccc}
      \toprule
                           & $u$-type & $d$-type & leptons  & $Q$ & $u_R$ & $d_R$ & $L$ & $l_R$ \\
      \midrule
      type I               & $\Phi_2$ & $\Phi_2$ & $\Phi_2$ & +   & $-$   & $-$   & +   & $-$   \\
      type II              & $\Phi_2$ & $\Phi_1$ & $\Phi_1$ & +   & $-$   & +     & +   & $-$   \\
      flipped (FL)         & $\Phi_2$ & $\Phi_1$ & $\Phi_2$ & +   & $-$   & $-$   & +   & +     \\
      lepton-specific (LS) & $\Phi_2$ & $\Phi_2$ & $\Phi_1$ & +   & $-$   & +     & +   & $-$
      \\ \bottomrule
    \end{tabular}
    \caption{Four left columns: Definition of the four Yukawa types of the
      $\mathbb{Z}_2$-symmetric 2HDM, given by the way each Higgs
      doublet couples to the various fermion types. Five right columns:
      Corresponding $\mathbb{Z}_2$
      assignment for the quark $Q$ and lepton $L$ doublets, the up-type
      quark singlet $u_R$, the down-type quark singlet $d_R$ and the
      lepton singlet $l_R$. \label{tab:yuycoup}}
  \end{center}
\end{table}

Altogether, the CN2HDM is then defined by 14 independent input
parameters which we choose to be
\begin{align}
  \alpha_{1-6},\, \tan \beta,\, m_{H_a},\, m_{H_b},\, m_{H_D},\,
  m_{H^\pm},\, v_s,\, \Re(m_{12}^2), \,
  \text{Type}(\text{2HDM})\,, \nn
\end{align}
where $m_{H_D}$ denotes the mass of the Dark Matter candidate $H_D
  \equiv a$ given by $m_{H_D} \equiv m_{\text{DM}}$. The masses
$m_{H_a},$ $m_{H_b}$ are any of the four Higgs mass eigenstates. One
of them is set to be equal to measured Higgs mass value of 125.09\gev
\cite{Aad:2015zhl}.
Note that the mass matrix components $(\mathcal{M}_N^2)_{\rho_1
      \rho_3}$ and $(\mathcal{M}_N^2)_{\rho_2 \rho_3}$ are not independent
and, in addition, the $(\mathcal{M}_N^2)_{s \rho_3}$ component is
vanishing. This leads to two dependent neutral masses
$m_{H_c}$, $m_{H_d}$ in the CN2HDM model,
\begin{align}
  \begin{split}
    m_{H_c}^2 &= \frac{m_{H_a}^2  \Big(R_{a4}  (R_{a3}  R_{c1} - R_{a1}  R_{c3}) -
    \tan\beta( R_{a3}  R_{a4}  R_{c2}  -
    R_{a2}  R_{a4}  R_{c3} ) \Big)}{
    R_{d4}  (R_{c3}  R_{d1} - R_{c1}  R_{d3} +
    (R_{c2}  R_{d3} -R_{c3}  R_{d2})  \tan\beta)}  \label{eq:mh3} \\
    &+
    \frac{m_{H_b}^2  \Big(R_{b4}  (R_{b3}  R_{c1} - R_{b1}  R_{c3}) -
    \tan\beta ( R_{b3}  R_{b4}  R_{c2}  -
    R_{b2} R_{b4}  R_{c3}) \Big)}{
    R_{d4}  (R_{c3}  R_{d1} - R_{c1}  R_{d3} +
    (R_{c2}  R_{d3} -R_{c3}  R_{d2})  \tan\beta)}
  \end{split}
\end{align}
\begin{align}
  \begin{split}
    m_{H_d}^2 &=  \frac{m_{H_a}^2
    (-R_{a3}  R_{a4}  R_{d1} + R_{a1}  R_{a4}  R_{d3} +
    R_{a4}  (R_{a3}  R_{d2} - R_{a2}  R_{d3})  \tan\beta)}{
    R_{c4}  (R_{c3}  R_{d1} - R_{c1}  R_{d3} +
    (R_{c2}  R_{d3} - R_{c3}  R_{d2})  \tan\beta)}    \label{eq:mh4} \\
    &+
    \frac{m_{H_b}^2
    (-R_{b3}  R_{b4}  R_{d1} + R_{b1}  R_{b4}  R_{d3} +
    R_{b4}  (R_{b3}  R_{d2} - R_{b2}  R_{d3})  \tan\beta)}{
    R_{c4}  (R_{c3}  R_{d1} - R_{c1}  R_{d3} +
    (R_{c2}  R_{d3} - R_{c3}  R_{d2})  \tan\beta)}\,.
  \end{split}
\end{align}
After computing these mass values, the four neutral Higgs mass eigenstates
$H_{a,b,c,d}$ are ordered in mass and renamed as $H_{1,2,3,4}$ such
that $m_{H_1} \le m_{H_2} \le m_{H_3} \le m_{H_4}$. The mixing matrix
elements are reordered and renamed accordingly.

The Feynman rule for the Higgs boson couplings to the massive gauge
bosons, $V = W, Z$, is given by ($i=1,2,3,4$)
\begin{align}
  i g_{\mu\nu} c(H_i VV) g_{h_{SM} VV} \,,
\end{align}
where $g_{h_{SM} VV}$ is the coupling of the SM Higgs boson to the
gauge bosons, and $c(H_i VV)$ is the effective coupling of the
CN2HDM Higgs bosons $H_i$ to the gauge bosons,
\begin{align}
  c(H_i VV) = & \cos\beta R_{i1} + \sin\beta R_{i2} \,.
\end{align}
The Yukawa Lagrangian has the form
\begin{align}
  \mathcal{L}_Y = - \sum_{i=1}^3 \frac{m_f}{v} \bar{\psi_f}\left( c^e(H_i ff) + i c^o (H_i ff) \gamma_5 \right) \psi_f H_i\,,
\end{align}
and the coefficients of the CP-even and -odd Yukawa couplings $c^e$
and $c^o$, respectively, are presented in Tab.~\ref{CN2HDM:CouplingsFermions}.
\begin{table}
  \centering
  \begin{tabular}{ccccccccc}
    \toprule
                     & \multicolumn{2}{c}{Type I} & \multicolumn{2}{|c}{Type II} & \multicolumn{2}{|c}{LS}      & \multicolumn{2}{|c}{FL}                                                                                                                                \\
                     & $c^e_i$                    & $c^o_i$                      & \multicolumn{1}{|c}{$c^e_i$} & $c^o_i$                      & \multicolumn{1}{|c}{$c^e_i$} & $c^o_i$                     & \multicolumn{1}{|c}{$c^e_i$} & $c^o_i$                     \\ \midrule
    up-type quarks   & $\frac{R_{i2}}{\sin\beta}$ & $- \frac{R_{i4}}{\tan\beta}$ & $\frac{R_{i2}}{\sin\beta}$   & $- \frac{R_{i4}}{\tan\beta}$ & $\frac{R_{i2}}{\sin\beta}$   & $-\frac{R_{i4}}{\tan\beta}$ & $\frac{R_{i2}}{\sin\beta}$   & $-\frac{R_{i4}}{\tan\beta}$ \\
    down-type quarks & $\frac{R_{i2}}{\sin\beta}$ & $\frac{R_{i4}}{\tan\beta}$   & $\frac{R_{i1}}{\cos\beta} $  & $-R_{i4} \tan\beta$          & $\frac{R_{i1}}{\cos\beta}$   & $R_{i4} \tan\beta$          & $-\frac{R_{i4}}{\tan\beta}$  & $\frac{R_{i2}}{\sin\beta}$  \\
    leptons          & $\frac{R_{i2}}{\sin\beta}$ & $\frac{R_{i4}}{\tan\beta}$   & $\frac{R_{i1}}{\cos\beta}$   & $-R_{i4}\tan\beta$           & $\frac{R_{i2}}{\sin\beta}$   & $-\frac{R_{i4}}{\tan\beta}$ & $R_{i4} \tan\beta$           & $\frac{R_{i1}}{\cos\beta}$  \\ \bottomrule
  \end{tabular}
  \caption{The couplings between two fermions and a neutral Higgs boson $H_i$ of the form $c^e(H_i ff) + i  c^o (H_i ff) \gamma_5$, with $c^{e,o}_i \equiv c^{e,o}(H_i ff)$.}
  \label{CN2HDM:CouplingsFermions}
\end{table}
In this work we restrict ourselves to the discussion of the CN2HDM type I and type II.

\section{Numerical Analysis}\label{sec:numericalanalysis}
For our numerical analysis we only take into account parameter points
that are allowed by current experimental and theoretical
constraints which we obtain from a scan in the CN2HDM parameter
space. To perform the scan and apply the constraints, the CN2HDM was
implemented in the {\tt C++} code, \texttt{ScannerS}
\cite{Coimbra:2013qq,Ferreira:2014dya,Muhlleitner:2016mzt,Muhlleitner:2020wwk},
and interfaced to {\tt CN2HDM\_HDECAY} via the interface
\texttt{AnyHDecay}~\cite{AnyHDecayLink}. The Fortran code {\tt
    CN2HDM\_HDECAY} computes the CN2HDM Higgs decay widths and branching
ratios including state-of-the-art higher-order corrections based
on {\tt HDECAY 6.51}~\cite{Djouadi:1997yw,Djouadi:2018xqq}.\footnote{The code {\tt
      CN2HDM\_HDECAY} builds up on the program {\tt N2HDECAY}~\cite{Muhlleitner:2016mzt,Engeln:2018mbg} that is
  the extension of {\tt HDECAY} to the N2HDM. For electroweak
  corrections to the N2HDM decay widths, see
  Ref.~\cite{Krause:2017mal,Krause:2019oar}.} The program can be downloaded
from the url:
\begin{center}
  \href{https://www.itp.kit.edu/~maggie/CN2HDM_HDECAY}{\tt https://www.itp.kit.edu/$\sim$maggie/CN2HDM\_HDECAY} \,.
\end{center}

On the theoretical side {\tt ScannerS} checks whether perturbative
unitarity holds, whether the potential is bounded from below and if
the chosen vacuum is the global minimum. The former two conditions
can be derived from the literature. We ensure tree-level perturbative
unitarity by requiring the eigenvalues of the $2 \to 2$ scalar scatter
matrix to be below an absolute upper value given by $8\pi$~\cite{Horejsi:2005da}. The required formulae can easily be generated with the script that is
shipped together with {\tt ScannerS}.
We apply the requirement of
boundedness from below of the potential in the strict sense by
requiring it to be strictly positive as the fields approach
infinity. We translated the necessary and sufficient conditions given
in Ref.~\cite{Klimenko:1984qx} to our notation.

On the experimental side the model has to comply with the LHC Higgs
data. Valid Higgs spectra must hence contain a Higgs boson with a mass
of 125.09 GeV~\cite{Aad:2015zhl} that behaves SM-like, \ie is compatible
with the measured Higgs data, and will be denoted $h_{125}$ from now
on. Agreement of the $h_{125}$ signal rates with the observations at
the $2\sigma$ level is checked by {\tt
    HiggsSignals-2.5.1}~\cite{Bechtle:2013xfa, Bechtle:2020uwn} that is linked to {\tt ScannerS}. Through the
link to \texttt{HiggsBounds-5.9.0}~\cite{Bechtle:2008jh,Bechtle:2011sb, Bechtle:2013wla, Bechtle:2015pma, Bechtle:2020pkv} the exclusion bounds from searches for extra scalars
are taken into account.

Flavour constraints from $B$ observables are mainly sensitive to the
charged Higgs sector, which, in the CN2HDM, remains unchanged from the
2HDM\@. We can therefore reuse existing 2HDM $2\sigma$ exclusion
bounds in the $m_{H^\pm}-\tan\beta$ plane~\cite{Haller:2018nnx}. This
leads to a lower bound of $m_{H^\pm} >
  580$~GeV, dominated by $B\to X_s\gamma$
measurements~\cite{charged580}, in the type II and lepton-specific
(CN)2HDM\@. For type I Yukawa sectors, the bound is much weaker and
more strongly dependent on $\tan\beta$.
The model must comply with electroweak precision
measurements and we demand $2\sigma$ compatibility of the $S,$ $T,$
and $U$ parameters computed with the general formulae given in \cite{STU1,
  STU2} with the SM fit \cite{Haller:2018nnx} including the full
correlations.
Since we allow for CP violation the model also has to comply with the
constraints from the measurements of the electric dipole moments. The
most stringent limit is the upper bound on the electric dipole
moment of the electron by the ACME collaboration
\cite{Andreev:2018ayy}.
Finally, we check for the DM constraints via an implementation of the
CN2HDM in \texttt{Micromegas-5.2.7}~\cite{ Belanger:2006is, Belanger:2008sj,
  Belanger:2010gh,Belanger:2013oya,Belanger:2014vza,Barducci:2016pcb,Belanger:2018ccd,Belanger:2020gnr}, which calculates the DM relic density and the
direct and indirect detection cross sections.


For our parameter scan we identify $H_a\equiv h_{125}$ and fix $m_{H_a}=125.09\,\text{GeV}$. The remaining input parameters are varied within the
following ranges,
\begin{equation}
  \begin{aligned}
                                       & -\frac{\pi}{2}\leq \alpha_{1..6} \leq \frac{\pi}{2}\,, &                             & 0.5 \leq  t_\beta \leq 25\,, \\[0.1cm]
                                       & 50\, \text{GeV} \leq m_{H_b},
    m_{H^\pm} \leq 1.5\, \text{TeV}\,, &                                                        & 1 \,\text{GeV} \leq m_{H_D}
    \leq 1.5 \,\text{TeV}\,,\vphantom{\frac{\pi}{2}}                                                                                                         \\
                                       & 1\, \text{GeV} \leq v_s \leq 3000\, \text{GeV} \,,     &                             & 10^{-3} \, \text{GeV}^2 \leq
    \text{Re}(m_{12}^2) \leq\,  5\times 10^5 \, \text{GeV}^2\,.\vphantom{\frac{\pi}{2}}
  \end{aligned}
\end{equation}
Note, that in our scan we exclude points of the parameter space where the
discovered Higgs signal is built up by two nearly degenerate Higgs
boson states by forcing the non-SM scalar masses to be outside the
mass window $m_{h_{125}}\pm 5$~GeV.

\section{Impact of the DM Observables \label{sec:DMconstraints}}

\begin{figure}
  \centering
  \includegraphics[width=0.9\textwidth]{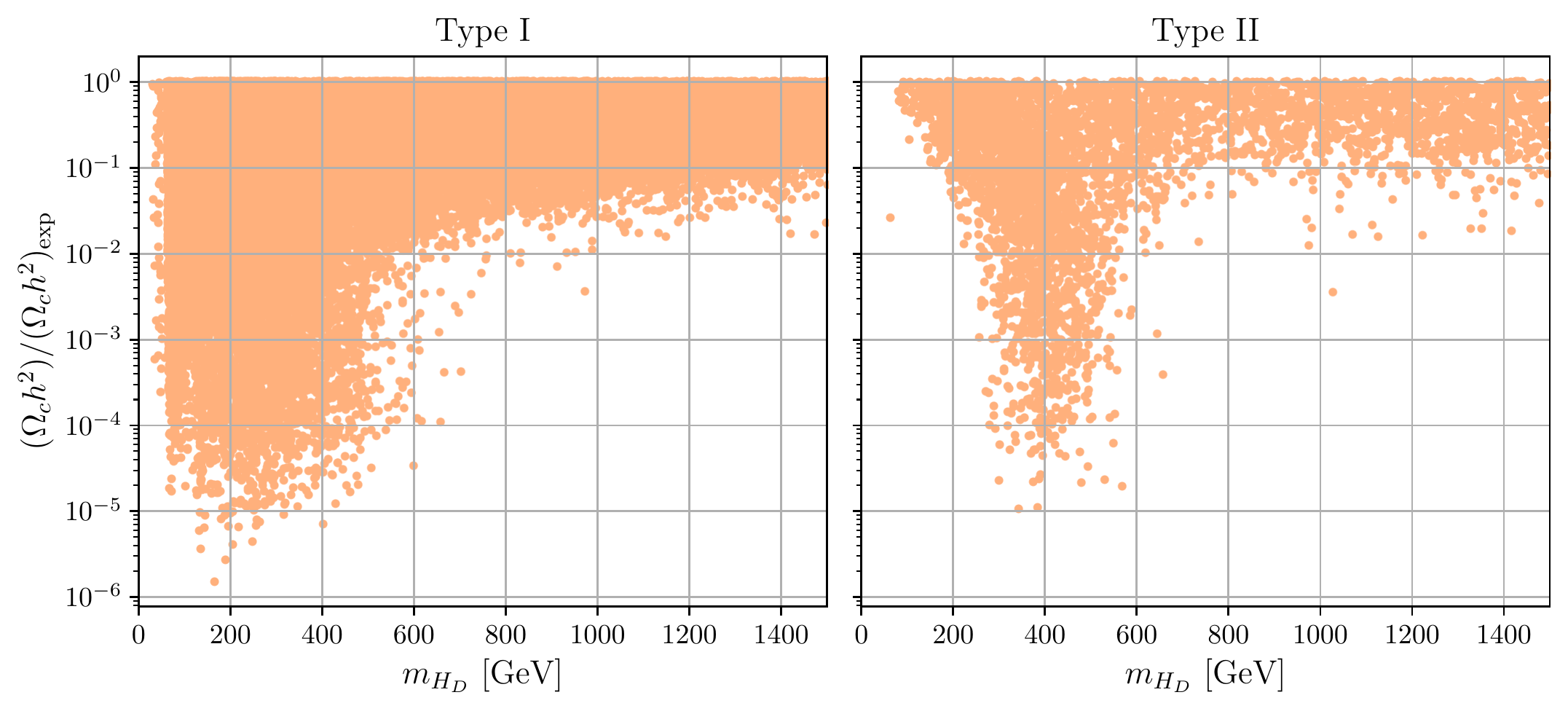}
  \caption{CN2HDM relic density normalised to the experimental
    value for type I (left) and type II (right) plotted against the DM
    mass $m_{H_D}$. }
  \label{fig:FracDMRelicDensity}
\end{figure}

We start
our phenomenological analysis of the model by investigating its
interplay with DM observables. The most important DM observable is
the DM relic density measured by the
space telescope Planck~\cite{Planck:2018vyg}. We require valid
parameter points to have a relic density that is under-abundant \ie
the predicted DM relic density is
below or equal to the observed relic density by Planck,
\begin{equation}
  (\Omega_c h^2)_{\text{exp}} = 0.1200 \pm 0.0012 \,. \label{eq:DMRD}
\end{equation}
Figure~\ref{fig:FracDMRelicDensity} depicts the obtained DM relic
density values in the CN2HDM type I (left) and type II (right)
normalised to the experimentally
measured value versus the DM mass. The plot
shows that in either Yukawa type the CN2HDM can account for the full
relic abundance for all allowed $m_{H_D}$.
In type I the lightest allowed $m_{H_D}$ are as low as
$m_{H_D}=30\gev$ with the highest allowed masses limited only by our scan range.
In type II $m_{H_D}$ values below $62.5\gev$ are exceedingly rare due
to overall heavier mass spectra enforced by the interplay of flavour
constraints and electroweak precision measurements.\footnote{In our
  scan, we found only one point where $62.5 \mbox{ GeV} > m_{H_D}$. A more
  dedicated scan in the low DM mass region might find more allowed
  points. Still, these scenarios are very
  rare.} However, for any values
of $m_{H_D}$ above that the type II CN2HDM can saturate the observed
relic density as well. The overall heavier mass spectrum in type II is
also the reason for the generally larger relic densities at low
$m_{H_D}$ in type II compared to type I. Due to the large mass
difference between the DM particle and the visible sector scalars that
mediate its interactions the 2-to-2 scattering
processes that deplete the relic density are less
efficient.

\begin{figure}
  \centering
  \includegraphics[width=0.9\textwidth]{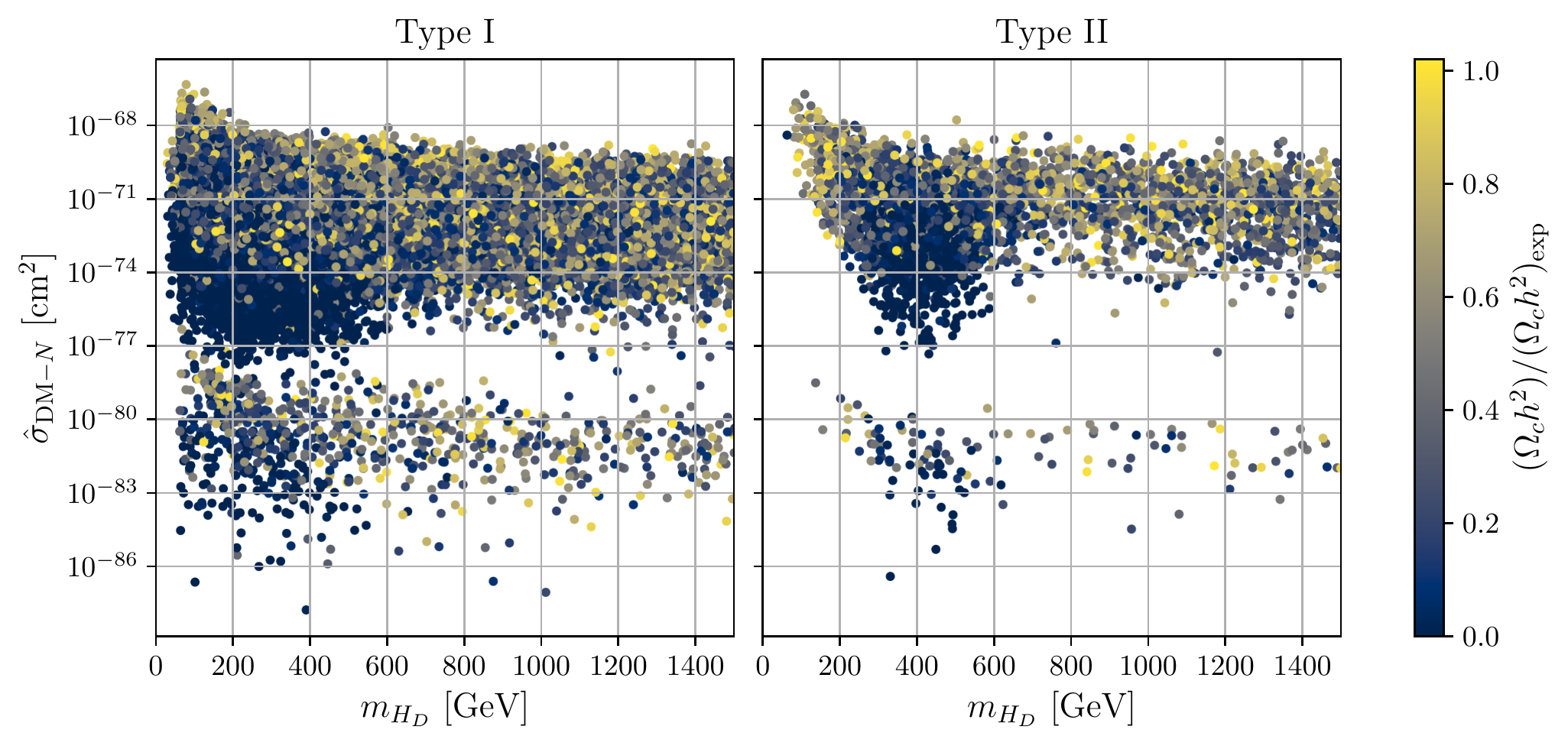}
  \caption{Effective SI direct detection DM-nucleon cross section
  $\hat{\sigma}_{\text{DM-N}}$ in cm$^2$ as a function of the DM mass
  $m_{H_D}$ for the CN2HDM type I (left) and type II
  (right).}\label{Fig:NucleonDMXS}
\end{figure}

Direct searches for DM particles have been and are being performed by
various experimental collaborations. The current strongest bound on
the spin-independent (SI) DM-nucleon scattering cross section ---
which is the relevant quantity for scalar mediators --- was obtained
by the \texttt{XENON1T}~\cite{XENON:2018voc} experiment. Since the
direct detection limits are derived assuming a DM relic density
equivalent to the central
experimental value of \cref{eq:DMRD}, they have to be
rescaled to the relic density values predicted in our model as
\begin{align}
  \hat{\sigma}_{\text{DM-N}} = \sigma_{\text{DM-N}} \frac{\Omega_c
    h^2}{(\Omega_c h^2)_{\text{exp}} }\,.
\end{align}
The resulting effective spin-independent (SI) DM-nucleon cross-section
$\hat\sigma_\text{DM-N}$ is plotted against the
DM mass in \cref{Fig:NucleonDMXS} (left) for type I
and (right) for type II\@. The colour code denotes the fraction of the observed relic
density predicted by our model.

\Cref{Fig:NucleonDMXS} demonstrates that for both CN2HDM types
the effective SI DM-nucleon cross section is heavily suppressed. All
of the points are well below the
neutrino-floor and hence are not in reach of direct
detections experiments. The cross sections are suppressed by more than
20 orders of magnitude across the whole DM mass range
compared to the upper limits on $\sigma_{SI}$ by
\texttt{XENON1T}~\cite{XENON:2018voc}. This is the consequence of a
cancellation mechanism present in the CN2HDM, which makes the
tree-level SI DM-nucleon cross section vanish in the limit of
vanishing momentum transfer. The non-zero SI cross section values
shown in \cref{Fig:NucleonDMXS} are due to higher-order QCD corrections
taken into account by {\tt MicrOmegas} which remain tiny due to the
loop suppression. A simpler model realising this
mechanism is discussed in Ref.~\cite{Gross:2017dan}, with electroweak
corrections relevant for the SI cross section in this model discussed in
Refs.~\cite{Azevedo:2018exj,Ishiwata:2018sdi,Glaus:2020ihj}. This
means that the CN2HDM can easily account for 100\% of the observed DM
density while remaining entirely out of reach of direct DM detection
experiments.

\begin{figure}[ht!]
  \centering
  \includegraphics[width=0.6\textwidth]{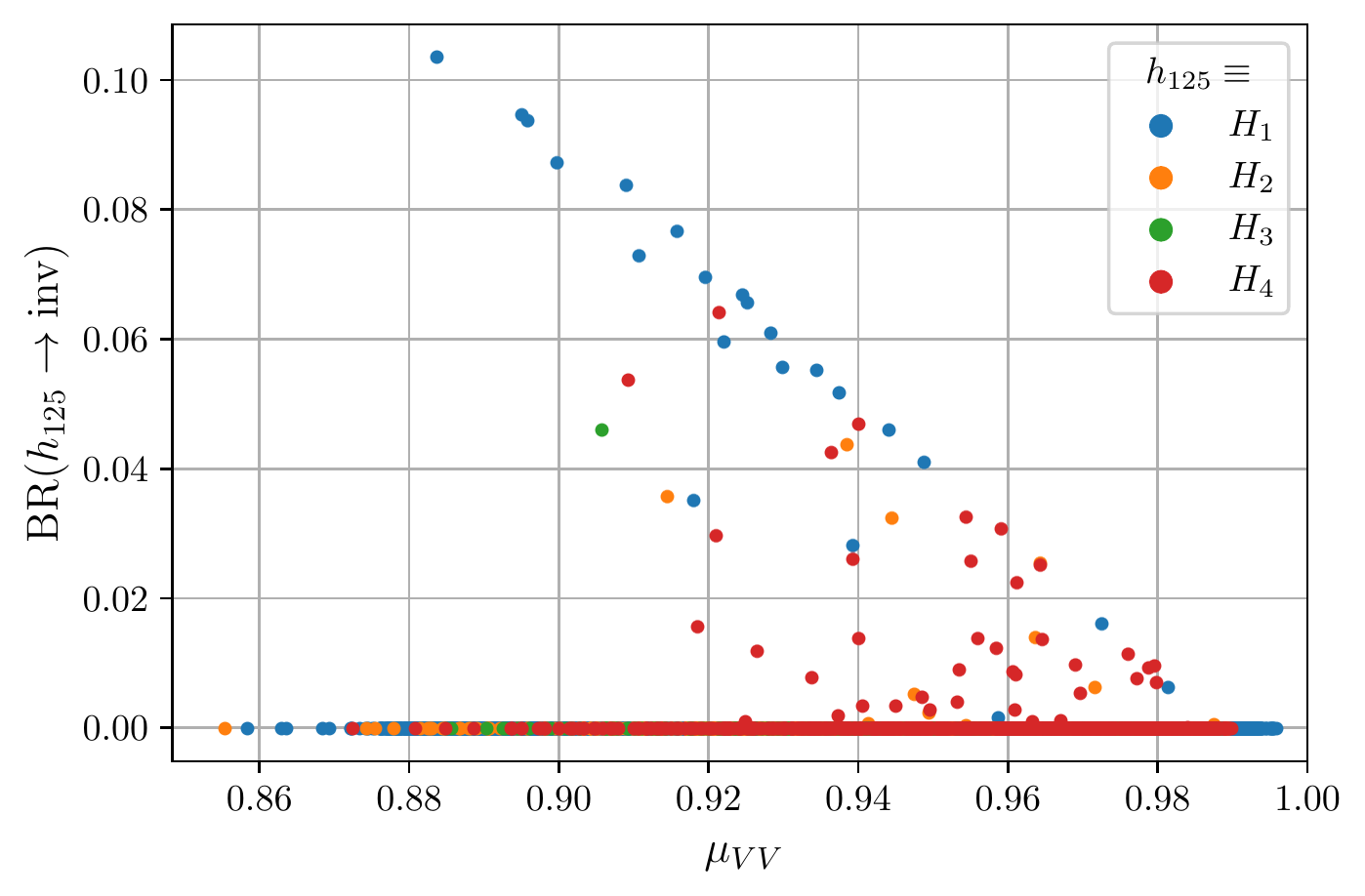}
  \caption{Type I: The branching ratio of the $h_{125}$ decaying into
    a pair of DM particles, for all types of $h_{125}$ mass ordering, plotted
    against the signal rate $\mu_{VV}$ into a massive gauge boson pair
    ($V\equiv Z,W^\pm$).}
  \label{Fig:BRh125Inv}
\end{figure}

Another way to investigate the dark sector is through invisible decays
of particles at the collider.
Current experimental measurements of
the $h_{125}$ Higgs decaying to invisible particles place an
exclusion limit on the Higgs $\to$ invisible branching ratio of 0.11
at 95\% C.L.~\cite{ATLAS:2019cid}. In \cref{Fig:BRh125Inv} we show
the invisible branching ratios obtained in the type I CN2HDM for all parameter
points passing our constraints as a function of the SM-normalised signal rate
$\mu_{VV}$ in the massive gauge boson final states $V\equiv Z,$
$W^\pm$. The colour code
indicates the different $h_{125}$ scenarios, namely for the blue
points $h_{125}$ is the lightest of the neutral non-DM Higgs bosons, $H_1$,
for the yellow points it is $h_{125}\equiv H_2$,
the green ones refer to $h_{125}\equiv H_3$ and the red points
denote scenarios where the heaviest Higgs boson, $H_4$, is
$h_{125}$. The $\mu_{VV}$ rate cannot be
smaller than 0.855 in any scenario. The figure shows
that for all allowed $\mu_{VV}$ values the branching ratio remains
below the direct experimental upper limit of 0.11 so that the Higgs rate
measurements in SM final states are still the most limiting
constraints on the invisible decay width of $h_{125}$. In the CN2HDM,
however, the direct limit from searches for $h_{125}\to\text{invisible}$ is
not far from being competitive. For future conclusive
statements it will therefore become necessary to include higher-order
corrections in the computation of the related branching ratio. We found a
similar behaviour as the one shown in \cref{Fig:BRh125Inv} for the
dark phases of the N2HDM~\cite{Engeln:2018mbg}, where the Higgs-to-invisible
branching ratio remains below 10\% after applying all Higgs constraints.

\begin{figure}[h!]
  \centerline{\includegraphics[width=.9\textwidth]{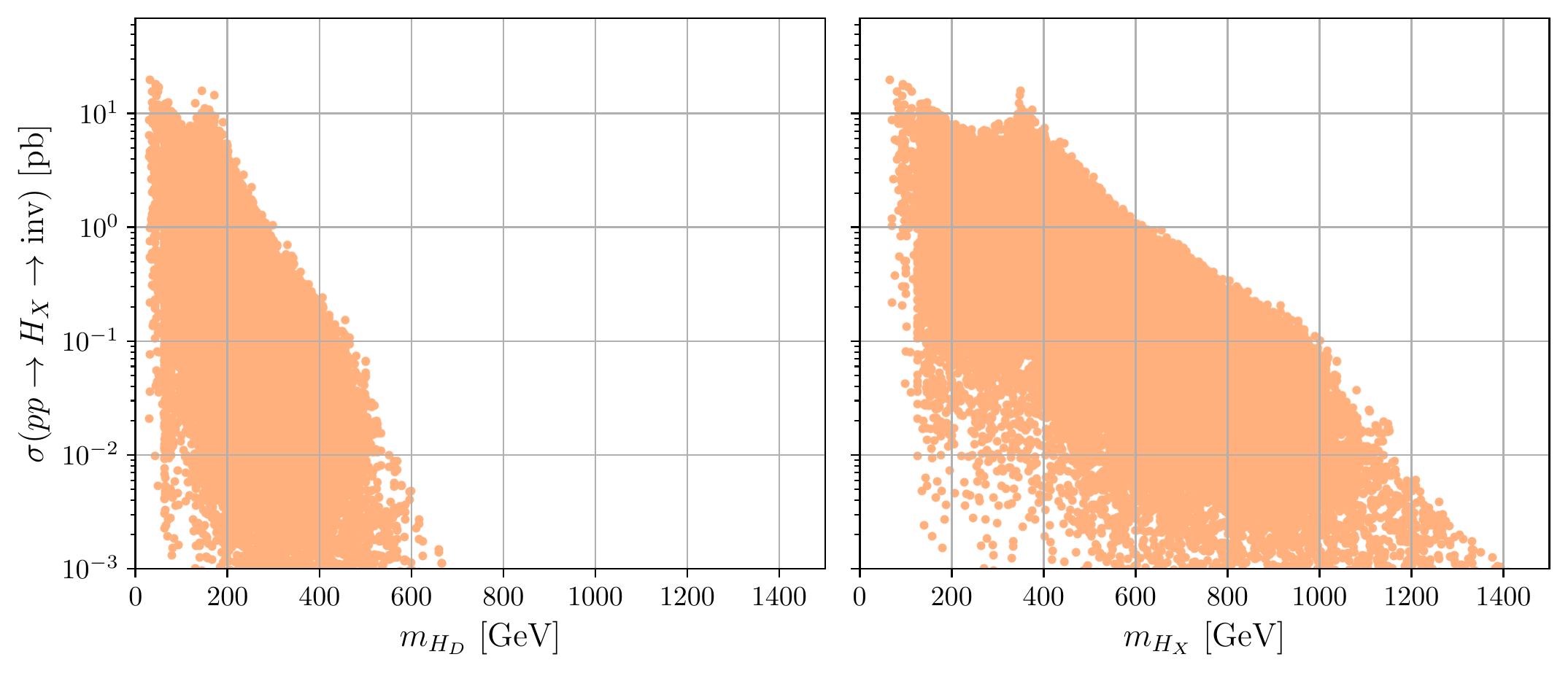}}
  \caption{Type I: The production rate of all visible neutral Higgs
    bosons $H_X$ ($X=1,2,3,4$) with subsequent decay
    into invisible particles as function of the DM mass $m_{H_D}$ (left), and as
    function of the mass of the decaying Higgs boson $m_{H_X}$
    (right).}
  \label{Fig:RateHInv}
\end{figure}

In the CN2HDM type II, only $H_1$ can be the SM-like Higgs
$h_{125}$. All other scenarios where $h_{125}$ is not the lightest
neutral scalar in the visible sector
are already excluded by the present constraints. We find
$\text{BR}(h_{125}\to\text{invisible}<10\%)$ in the type II version of
the model as well. However, for the vast majority of the valid
parameter points in type II $m_{H_D}> 62.5\gev$ such that the
branching ratio is zero.

In \cref{Fig:RateHInv} we show for the CN2HDM type I the
production cross sections of all visible
neutral Higgs bosons, $H_X$ ($X=1,2,3,4$), \ie also the non-SM-like ones, times
their subsequent decay into invisible particles,
\begin{equation}
  \sigma (pp\to H_X \to \mbox{inv}) = \sigma(\text{ggF + bbH}) \times \mbox{BR}
  (H_X \to \mbox{inv}) \,,
\end{equation}
plotted against
$m_{H_D}$ (left) and against the mass $m_{H_X}$ of the decaying
particle (right).
All production cross sections have been computed at
$\sqrt{s}=13$~TeV with {\tt
    SusHi} v1.6.1~\cite{Harlander:2012pb,Harlander:2016hcx} and
are the sum of gluon production and associated production with a
bottom quark pair. For most of the points --- particularly those
  with the largest rates --- the invisible decay is the dominant decay mode
  of $H_X$, often with branching ratios above 90\%. In such a scenario a search in
  invisible final states may well be the only possibility to discover that particle.
As can be inferred from the plot, the maximum cross
section values at the lower end of the mass spectrum amount up to
25 pb which is reached for the case where $H_4=h_{125}$ in the decay
$H_2 \to$ invisible. We
observe a dip at $m_X \approx 250$~GeV. Here the heavier non-SM-like
Higgs bosons predominantly decay into a pair of SM-like $h_{125}$
Higgs bosons instead of a DM pair. This region corresponds to $m_{H_D}
  \approx 110$~GeV in the left plot. After increasing again to maximum
values around $m_{H_X}=350$~GeV where the gluon fusion production
cross section peaks, the rates quickly fall with increasing $H_X$
mass. Still, for heavy Higgs mass values of 1~TeV they amount up to
100~fb. They cross 1 fb at $m_{H_X}= 1.4$~TeV.

\begin{figure}
  \centering
  \includegraphics[width=.9\textwidth]{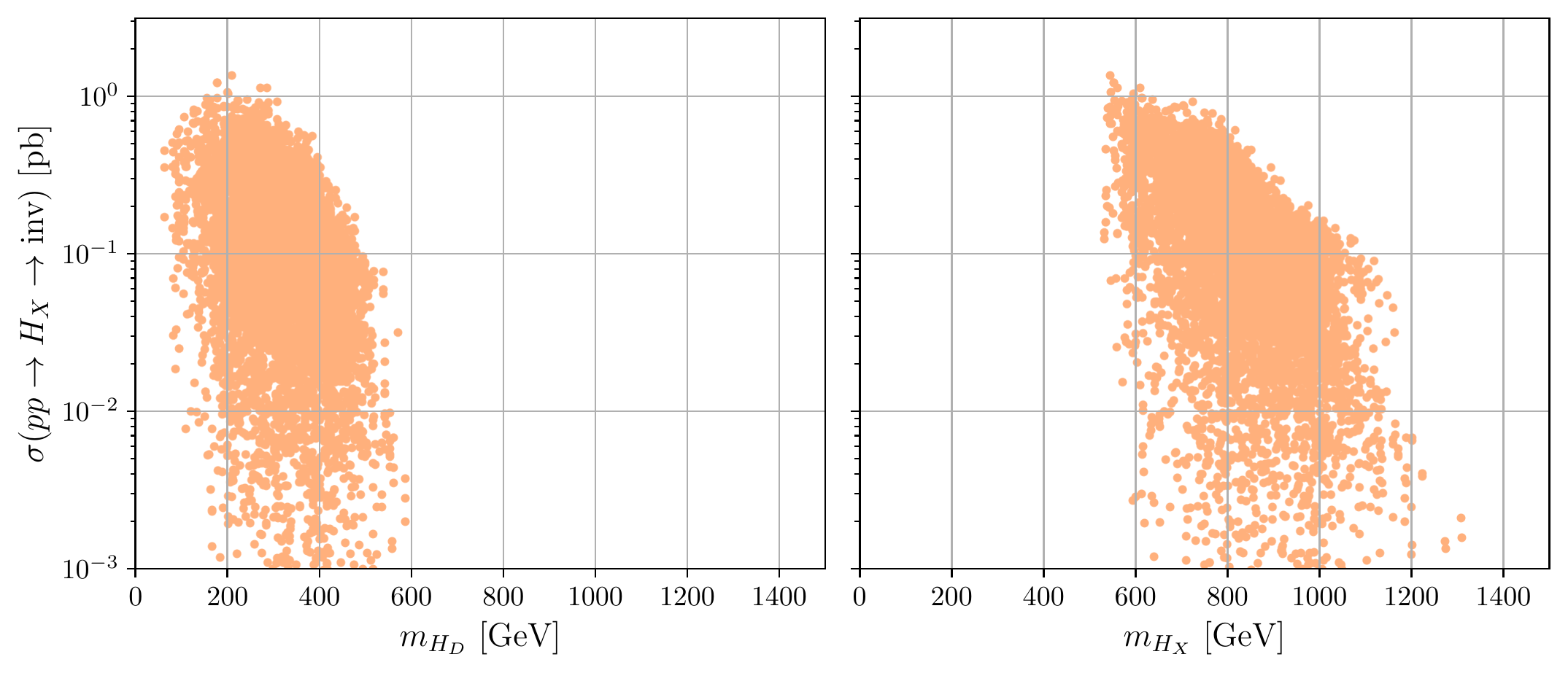}
  \caption{Type II: The production rate of all neutral Higgs bosons
    $H_X$ ($X=1,2,3,4$) with subsequent decay
    into invisible particles as function of the DM mass $m_{H_D}$ (left), and as
    function of the mass of the decaying Higgs boson $m_{H_X}$
    (right).}
  \label{Fig:RateHInv2}
\end{figure}

\Cref{Fig:RateHInv2} shows the same rates as \cref{Fig:RateHInv} but
for type II instead of type I Yukawa sectors. In type II,
the neutral non-SM like Higgs masses $H_{2,3,4}$ are overall pushed
to larger values, mainly due to the interplay of flavour and
electroweak precision constraints. These also exclude scenarios with
$m_{H_D} \le 62.5$~GeV apart from very rare cases as stated
above. With an overall heavier spectrum the rates are smaller
than in the type I CN2HDM and reach at most 1.3~pb.

\section{LHC Phenomenology of the CN2HDM Higgs Bosons \label{sec:higgspheno}}
We now want to investigate the collider phenomenology in
our model. The specific phenomenological feature of $h_{125}$ in our
model is the possibility of
non-vanishing singlet and/or CP-odd admixtures to the SM-like Higgs
mass eigenstate which will affect its couplings and hence its production and
decay rates. While we delay a detailed study of the phenomenology of
the non-SM-like Higgs bosons to a future work, we will also comment on
the overall mass spectra and their typical phenomenology.
We first discuss the phenomenology of a type I CN2HDM in some detail
and then comment on the differences in type II.

\subsection{The Type I Higgs Mass Spectrum}
If $h_{125}\equiv H_1$ the visible non-SM-like Higgs bosons can
decouple with all masses becoming large and similar. In this scenario,
the maximum allowed values of $m_{H_{2,3,4}},m_{H^\pm}$ are only
limited by our scan ranges. Decoupling is no longer possible if
$h_{125}$ is not the lightest of the neutral scalars and the Higgs
mass spectrum becomes increasingly lighter as $h_{125}\equiv H_{2,3,4}$. Thus for $H_2
  \equiv h_{125}$ the maximum values of $m_{H_{3,4}, H^\pm}$ are just
above 600~GeV. Moreover, we now have an additional light Higgs in
the spectrum with masses as low as $m_{H_1}=50$~GeV, where again the lower limits are
determined by the lower limits of our scan ranges. For the case with
$H_3 \equiv h_{125}$, $m_{H_4}$ reaches at most 415~GeV and
$m_{H^\pm}$ remains below 270~GeV. Additionally, we have two light
visible Higgs masses in the spectrum with masses below 125~GeV and
lower limits of $m_{H_1}=53$~GeV and $m_{H_2}=70$~GeV. For $H_4 \equiv
  h_{125}$ the heaviest visible Higgs boson is the charged one with a
maximum mass value of $m_{H^\pm}=190$~GeV. The three lighter Higgs
bosons below 125~GeV have lower limits of 55~GeV, 68~GeV and 82~GeV
for $m_{H_1}$, $m_{H_2}$ and $m_{H_3}$, respectively.

Let us give a brief generic overview of the signatures of the
non-SM-like Higgs bosons.
The main decay channels of the neutral Higgs bosons with masses below
125~GeV are those into lighter fermions and into a photon pair,
depending on the respective mass values. The neutral visible Higgs
bosons above 125~GeV mainly decay into top quark pairs, a neutral
Higgs plus $Z$ boson pair, a charged Higgs plus $W^\pm$ pair, or into
a pair of lighter neutral Higgs bosons. The specific decay channel
depends on the involved coupling and mass values.

The charged
Higgs boson decays, depending on the charged Higgs couplings
and mass values, mainly
into a top plus bottom quark pair, into a $\tau$ lepton and its
associated neutrino, or into a neutral Higgs plus $W^\pm$ pair.

\begin{table}
  \begin{center}
    \renewcommand{\arraystretch}{1.2}
    \begin{tabular}{ccccccc}
      \toprule
      $\alpha_1$      & $\alpha_2$                      & $\alpha_3$       & $\alpha_4$        & $\alpha_5$   &
      $\alpha_6$      & $\tan\beta$                                                                                            \\
      1.386           & -0.048                          & 1.263            & -0.028            & 1.395        & 0.0041 & 5.996 \\ \midrule
      $m_{H_a}$ [GeV] & $m_{H_b}$  [GeV]                & $m_{H_D}$  [GeV] & $m_{H^\pm}$ [GeV] & $v_s$  [GeV]
                      & $\mbox{Re}(m_{12}^2)$ [GeV$^2$] &                                                                      \\
      125.09          & 346.13                          & 1472             & 578.97            & 477.92       & 43713  &       \\ \bottomrule
    \end{tabular}\\[0.5cm]
    \begin{tabular}{ccccccc}
      \toprule
      $m_{H_3}$ [GeV] & $m_{H_4}$ [GeV]       & $\sigma_{\text{prod}} (H_4)$ [fb]
                      & BR($H_4 \to H_1 H_2$) & BR($H_2 \to H_1 H_1$)                             \\
      516.83          & 532.02                & 97.23                             & 0.463 & 0.577
      \\ \bottomrule
    \end{tabular}
    \caption{Type I: Input parameter values where
      $H_a \equiv H_1$ and $H_b \equiv H_2$ (top) and further
      information (bottom) on the $H_1H_1H_1$ benchmark point.}
    \label{tab:HiggsRateEx}
  \end{center}
\end{table}

A smoking gun signature for extended Higgs sectors with
more than two neutral visible Higgs bosons (\ie more than
in a 2HDM) would
be Higgs-to-Higgs cascade decays. In the case with $H_1
  \equiv h_{125}$ we can \eg find a triple SM-like Higgs $H_1 H_1 H_1$ production
signature at a rate of $\sigma_{\text{prod}} (H_4) \times
  \mbox{BR} (H_4 \to H_1 H_2) \times \mbox{BR} (H_2 \to H_1 H_1) = 26$~fb. The input
values for this benchmark point as well as additional information on
the remaining neutral Higgs mass values, the
$H_4$ production cross section and the involved Higgs-to-Higgs
branching ratios are given in \cref{tab:HiggsRateEx}.

\subsection{The Type I SM-Like Higgs Singlet and CP-odd Admixtures}
We define the singlet and CP-odd components of $h_{125}$ through the
respective elements in the mixing matrix $R$, namely,
\begin{equation}
  \begin{aligned}
    \mbox{singlet admixture:} & \quad R_{h_{125},3}^2\,, \\
    \mbox{CP-odd admixture:}  & \quad R_{h_{125},4}^2\,.
  \end{aligned}
\end{equation}
This means that the singlet (CP-odd) admixture is given by the matrix element
at the row associated with $h_{125}$ and the column
associated with $s$ ($\rho_3$).

\begin{table}
  \begin{center}
    \begin{tabular}{rcccc}
      \toprule
      $h_{125}\equiv$        & $H_1$ & $H_2$ & $H_3$ & $H_4$ \\ \midrule
      singlet admixture [\%] & 12.3  & 9.5   & 9.3   & 8.1   \\
      CP-odd admixture [\%]  & 6.2   & 7.1   & 6.0   & 2.9   \\ \bottomrule
    \end{tabular}
    \caption{Type I: The singlet and CP-odd admixture of the SM-like Higgs boson in \% for the four different cases of $h_{125}\equiv H_{1,2,3,4}$.}
    \label{tab:admixture}
  \end{center}
\end{table}

\begin{center}
  \begin{figure}
    \begin{center}
      \includegraphics[width=0.8\textwidth]{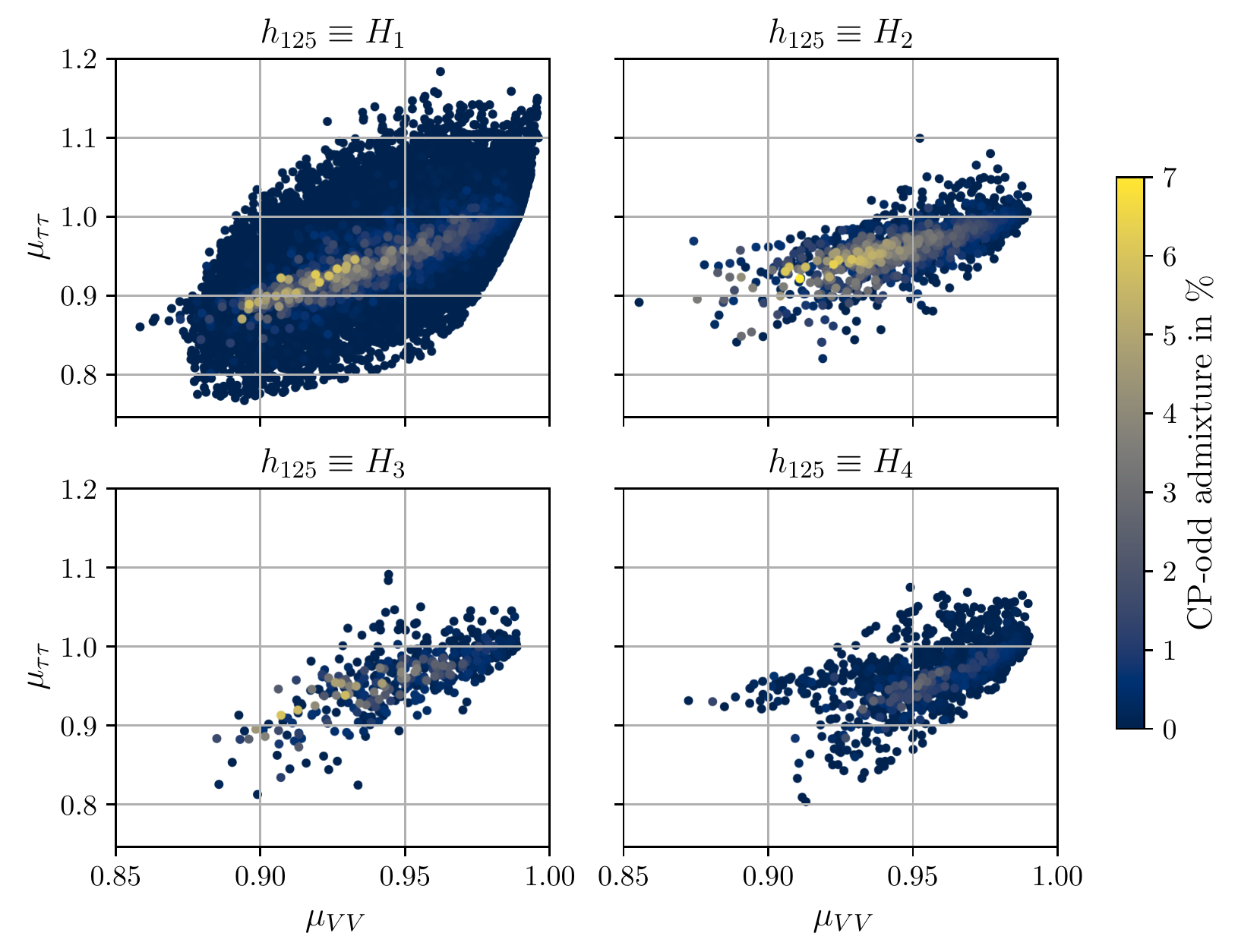}
    \end{center}
    \caption{Type I: The rate $\mu_{\tau\tau}$ versus $\mu_{VV}$ for
      $H_{1,2,3,4}$ being SM-like, respectively, with the colour code denoting
      the CP-odd admixture. }
    \label{fig:T1mu_Admix}
  \end{figure}
\end{center}

\begin{center}
  \begin{figure}
    \begin{center}
      \includegraphics[width=0.8\textwidth]{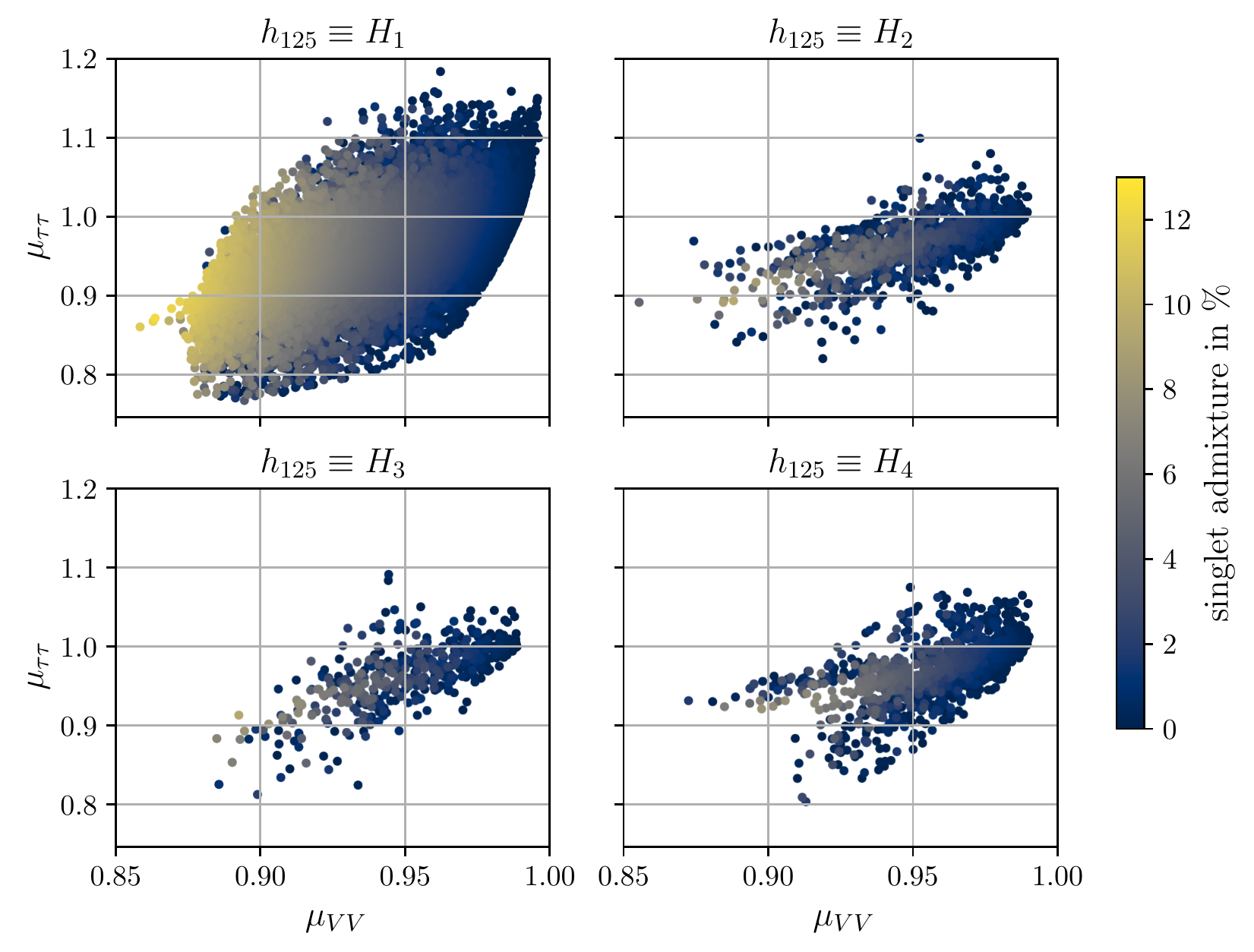}
    \end{center}
    \caption{Type I: The rate $\mu_{\tau\tau}$ versus $\mu_{VV}$ for
      $H_{1,2,3,4}$ being SM-like, respectively, with the colour code denoting
      the singlet admixture. }
    \label{fig:T1mu_singletAdmix}
  \end{figure}
\end{center}

In Table \ref{tab:admixture} we list the maximal separately reachable
singlet and CP-odd admixtures of $h_{125}$ for the four different
cases of $h_{125}\equiv H_{1,2,3,4}$ that are still in agreement with
all constraints. The largest singlet component (12.3\%) is reached for
$h_{125} = H_1$ and decreases for lighter Higgs mass spectra. The
maximum CP-odd admixture (7\%) is obtained for $h_{125} \equiv H_2$.
Figure~\ref{fig:T1mu_Admix} shows the impact of the CP-odd admixture
on the fermionic ($\mu_{\tau\tau}=\mu_{bb}$ in type I) and bosonic
signal strengths of $h_{125}$. We find that the CP-odd admixture decreases towards the
lower and upper limits of $\mu_{VV}$ and $\mu_{\tau\tau}$. The
maximum values are reached on the diagonal around $\mu_{\tau\tau} =
  \mu_{VV} \approx 0.92-0.94$. Figure~\ref{fig:T1mu_singletAdmix} shows
the corresponding plot for the singlet admixture. As the singlet
component has no couplings to any SM particles, the singlet admixture
decreases $\mu_{VV}$ and $\mu_{\tau\tau}$ simultaneously. This is
clearer for $\mu_{VV}$ since $c^2(h_{125}VV)<1$ and therefore the
coupling cannot be enlarged to counteract the singlet effects. For the
fermion coupling this is possible leading to points with substantial
singlet admixture also at larger $\mu_{\tau\tau}$.

\subsection{The Type I SM-Like Higgs Rates}
Since our model has an alignment limit, there are parameter sets where
all $h_{125}$ couplings to fermions and gauge bosons are very close to the
SM ones. We hence have to look for deviations in the Higgs rates from
the SM-case that are specific in our model. In
\cref{tab:rateranges} we summarise the ranges that are still
compatible with all constraints in
the production rates $\mu_{\tau\tau}=\mu_{bb}$, $\mu_{VV}$, and
$\mu_{\gamma\gamma}$ for the four different $h_{125}$
scenarios. As can be inferred from the table, the
maximum freedom is given for the case where $H_1 \equiv h_{125}$. In
all $h_{125}$ scenarios the $\mu_{\tau\tau}$ range can exceed the
SM-value (with the largest possible value of 1.18 obtained for $H_1$
being SM-like). The maximum Higgs rates into gauge bosons are $\le 1$
for $h_{125}\equiv H_{1,...,4}$. As for the
$\gamma\gamma$ rates, the lower limits in all four scenarios are
about the
same. However, the upper limit is largest for $H_1$ and equal to 1.14. In
the other three cases the upper $\mu_{\gamma\gamma}$ is 
$\le 1.04$. This is the result of the interplay between the trilinear Higgs
self-coupling $\lambda_{h_{\text{125}}H^+H^-}$ and the mass of the
charged Higgs boson running in the loop of the loop-induced Higgs coupling to the
photons. 

\begin{table}
  \begin{center}
    \begin{tabular}{lcccc}
      \toprule
                           & $H_1$       & $H_2$       & $H_3$       & $H_4$       \\ \midrule
      $\mu_{\tau\tau}$     & $0.77-1.18$ & $0.82-1.11$ & $0.81-1.09$ & $0.80-1.08$ \\
      $\mu_{VV}$           & $0.86-1.00$ & $0.86-0.99$ & $0.88-0.99$ & $0.87-0.99$ \\
      $\mu_{\gamma\gamma}$ & $0.83-1.14$ & $0.82-1.04$ & $0.82-1.01$ & $0.82-1.02$ \\ \bottomrule
    \end{tabular}
    \caption{Type I: The maximal and minimal signal rates of $h_{125}$
      compatible with  all applied constraints for the four different
      cases of $h_{125}\equiv H_{1,2,3,4}$. \label{tab:rateranges}}
  \end{center}
\end{table}

  \begin{figure}[ht!]
    \begin{center}
      \includegraphics[width=0.8\textwidth]{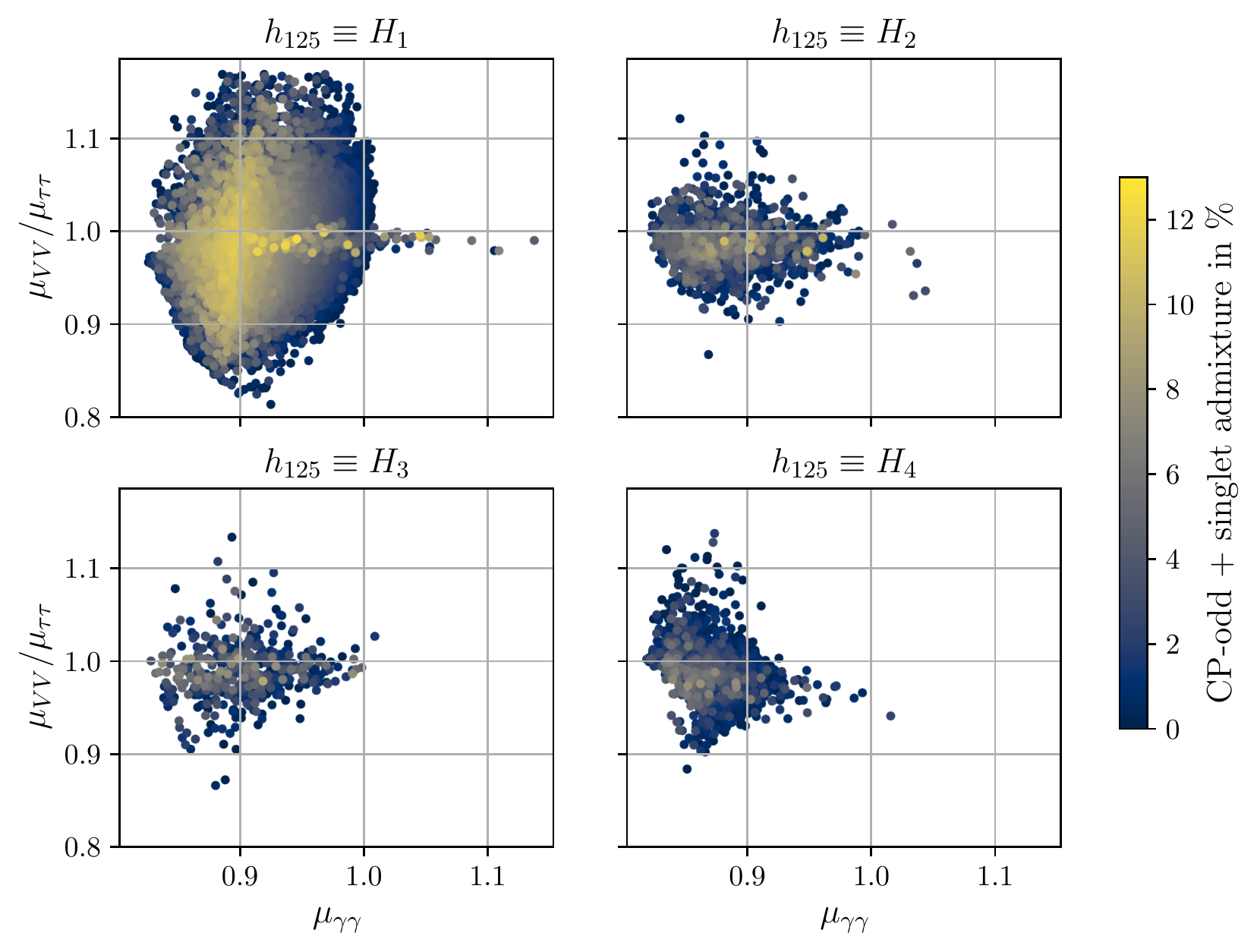}
    \end{center}
    \caption{Type I: The rates ratio $\mu_{VV}/\mu_{\tau\tau}$ versus $\mu_{\gamma
          \gamma}$ for $H_{1,2,3,4}$ being SM-like, respectively, with the colour code denoting
      the combined CP-odd and singlet admixture. }
    \label{fig:T1muGamGam_Admix}
  \end{figure}

\begin{figure}[h!]
    \centerline{\includegraphics[width=0.6\textwidth]{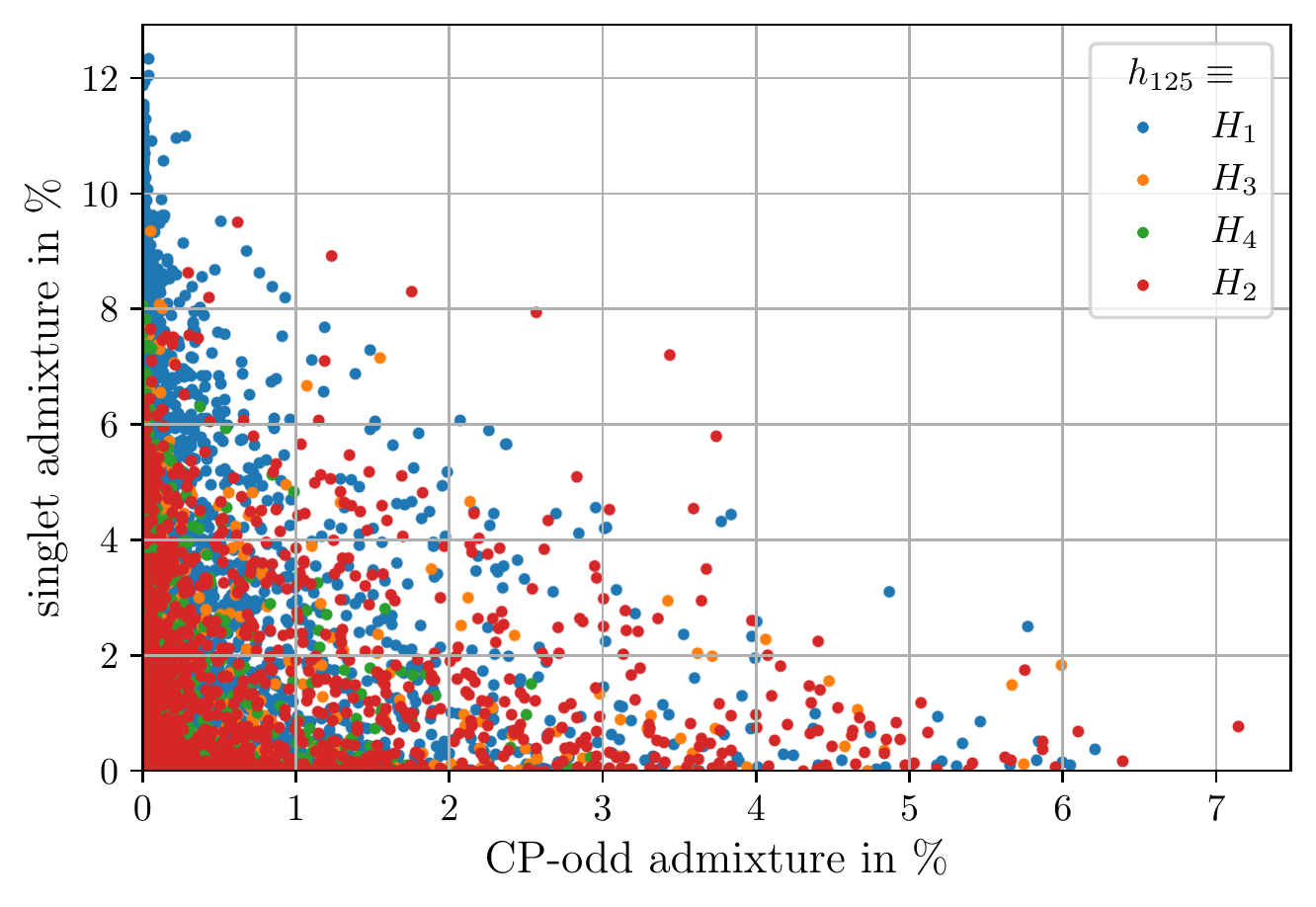}}
    \caption{Type I: The singlet admixture versus the CP-odd admixture in \% for
      $H_1\equiv h_{125}$ (blue), $H_2\equiv h_{125}$ (orange), $H_3\equiv
        h_{125}$ (green),  and $H_4\equiv h_{125}$
      (red).}
    \label{fig:T1SvCPV}
  \end{figure}

Figure~\ref{fig:T1muGamGam_Admix} shows the ratio $\mu_{VV}/\mu_{\tau
  \tau}$ plotted against $\mu_{\gamma\gamma}$, with 
the colour code denoting the combined CP-odd and singlet admixture of
the $h_{125}$ given by $R_{h_{125},3}^2+R_{h_{125},4}^2$.
The highest combined admixtures found in the four cases are $12.3\%$
when $h_{125} = H_1$, $10.9\%$ when $h_{125} = H_2$, $9.6\%$ when $h_{125}
  = H_3$ and $7.8\%$ when $h_{125} = H_4$. These values tend to be
dominated by the singlet admixture, and the
highest values of the singlet (CP-odd) admixture of the $h_{125}$ are
accompanied by low values of the CP-odd (singlet) admixture,  {\it
    cf.}~\cref{fig:T1SvCPV}.

We furthermore observe that the $\mu_{VV}$ and $\mu_{\tau\tau}$ values become
increasingly correlated with rising CP-odd plus singlet admixtures so
that their ratio is closer to 1 with larger
admixtures. The large mixing case is
hence barely distinguishable from the SM-case in this ratio. In the
$\mu_{\gamma\gamma}$ rate a faint trend to smaller admixtures can be
observed towards the edges of its allowed range. We furthermore
clearly observe that $\mu_{\gamma\gamma}$ ranges above 1 require
$\mu_{VV}/\mu_{\tau\tau}$ values close to 1 for $H_1 \equiv h_{125}$
whereas this is less pronounced in the other scenarios, which,
however, also have smaller maximal $\mu_{\gamma\gamma}$ values.

In summary, deviations in the individual $\mu_{\tau\tau}$ and
$\mu_{VV}$ rates correlate with non-zero CP-odd and/or singlet
admixtures, whereas their ratio does not allow us to make conclusive
statements about these admixtures. The latter can also be said about
the $\mu_{\gamma\gamma}$ rate. The photonic rates, however, allow us to
distinguish the $h_{125}$ scenarios. Values of above
1.04 point towards $H_1 \equiv h_{125}$. Furthermore,
$\mu_{\gamma\gamma}$ values above 1 
with $\mu_{VV}/\mu_{\tau\tau}$ deviating by more than 
about 0.2\% from the SM  ratio 1 may be a sign of $H_{2,3,4}$ being SM-like.

\subsection{The CN2HDM Type II}
In the CN2HDM type II, only points where the lightest visible neutral
Higgs boson is the $h_{125}$ survive the constraints. The highest combined
CP-odd-singlet admixture coincides with the highest singlet admixture
at $19.3\%$ and is higher than in type I. The CP-odd admixture can become as
large as 9\%. The reason for larger allowed admixtures is that in the
type II model the couplings to up- and to down-type fermions are
disentangled and we hence have more freedom in the production and
decay rates to accommodate larger non-SM admixtures to $h_{125}$
without violating the Higgs constraints.
The $\mu_{\tau\tau}$ range is somewhat smaller than in the type I case with
$H_1 \equiv h_{125}$, lying between 0.76 and 1.13. The
$\mu_{\tau\tau}$ rate is found to increase with increasing CP-odd admixture
as we have an additional CP-odd-like coupling contributing to the
Higgs-Yukawa interaction of the SM-like Higgs, and to decrease with
increasing singlet admixture.
The allowed $\mu_{VV}$ rate lies between 0.87 and 1.10 and thus clearly
exceeds the upper limit allowed in type I. An enhancement of
$\mu_{VV}$ is possible in type II since the coupling of $h_{125}$ to
down-type quarks --- and thus the dominant $h_{125}\to b\bar{b}$ decay
width and the total width --- can be reduced without simultaneously
reducing the $t\bar{t}$ coupling that dominates the gluon fusion
production cross section.
The allowed $\mu_{\gamma\gamma}$ range is nearly identical to the one of
$\mu_{VV}$, and neither of the two rates shows any sensitivity to the
non-SM-like admixtures.

\begin{figure}
  \centerline{\includegraphics[width=0.6\textwidth]{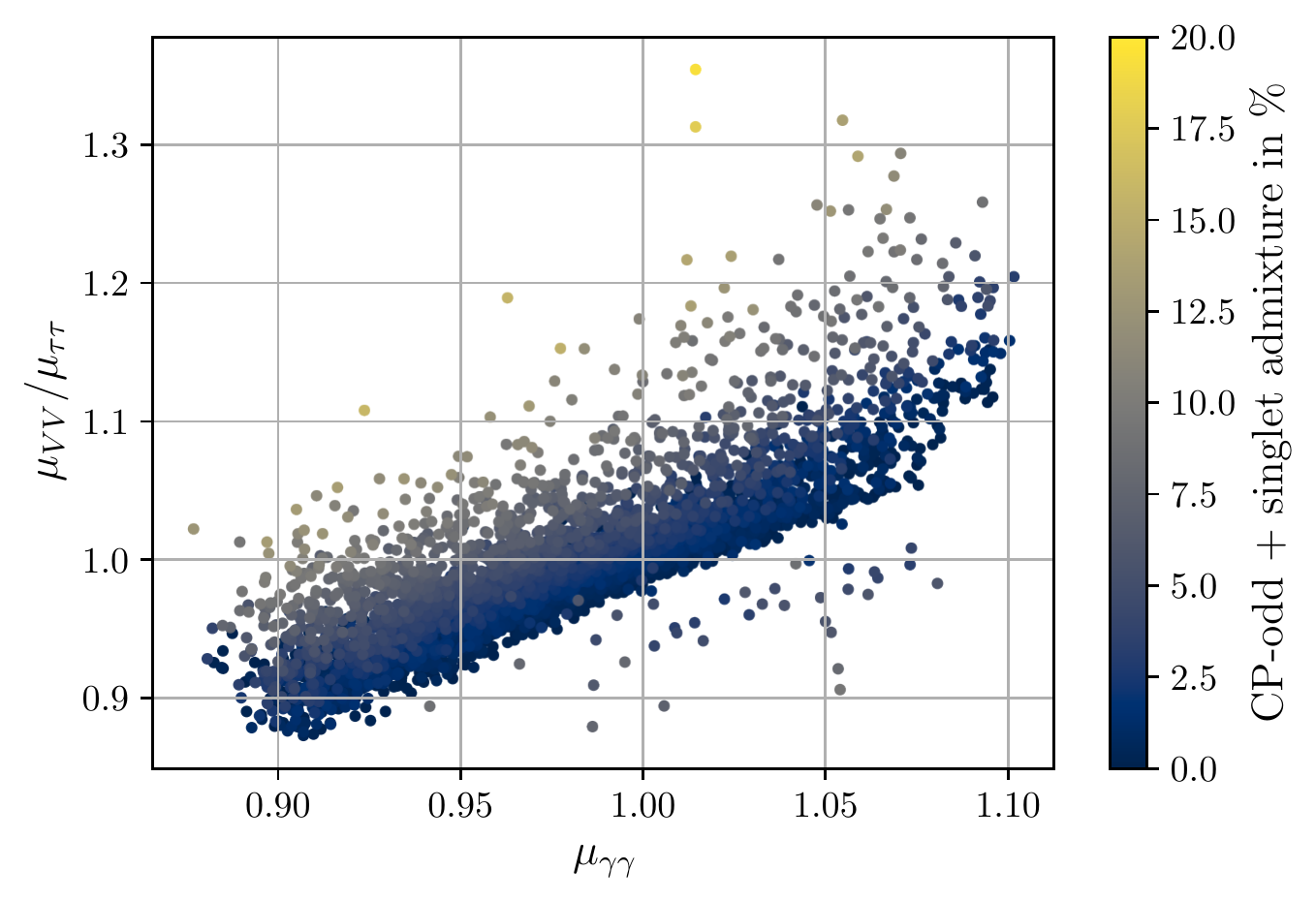}}
  \caption{Type II: The rates ratio $\mu_{VV}/\mu_{\tau\tau}$ versus $\mu_{\gamma
        \gamma}$ for $H_{1}$ being SM-like, the only case compatible with
    the constraints. The colour code denotes
    the combined CP-odd and singlet admixture. } 
  \label{fig:T2mu_Admix}
\end{figure}

In \cref{fig:T2mu_Admix} we depict the signal strength fraction
$\mu_{VV}/\mu_{\tau\tau}$ versus $\mu_{\gamma \gamma}$ for all type II points
passing our constraints. The colour code
indicates the combined CP-odd and singlet admixture.
The ratio $\mu_{VV}/\mu_{\tau\tau}$ can reach
values of up to $1.35$ which are significantly higher than in the CN2HDM type I.
At the same time, the lower bound of $0.87$ is also slightly higher than in type I. The ratio $\mu_{VV}/\mu_{\tau\tau}$
shows a dependence on the combined CP-odd
and singlet admixtures in contrast to type I, and increases with
increasing non-SM-like admixtures.

In summary, while we cannot use $\mu_{\gamma\gamma}$ to distinguish type I and type II,
we can use $\mu_{\tau\tau}$ and $\mu_{VV}$ and in particular their ratio. Furthermore, contrary to type I, the
ratio measurement of $\mu_{VV}$ and $\mu_{\tau\tau}$ allows for
conclusions on non-SM-like admixtures to the SM-like Higgs boson.

\section{Conclusions}\label{Conclusion}
In this work we have introduced the CN2HDM\@. It is based on the
CP-violating 2HDM extended by a complex singlet field that obeys a
discrete $\mathbb{Z}_2$ symmetry. Thereby we obtain a model that
implements all ingredients required to answer the most pressing
open questions of the SM\@: It contains a DM candidate and it features CP
violation, one of the three Sakharov conditions required for
successful baryogenesis. After working out the relevant equations, we have
implemented the model in {\tt ScannerS} and used that implementation to perform
scans in the parameter space of the model by simultaneously taking
into account the relevant theoretical and experimental
constraints. We investigated the type I and the type II
version of the model and analysed the interplay between DM observables
and the LHC phenomenology of the model. For the latter, we provided
the code {\tt CN2HDM\_HDECAY} for the computation of the CN2HDM Higgs
boson decay widths and branching ratios including
state-of-the-art higher-order corrections.

The Higgs spectrum of the CN2HDM consists of a stable DM candidate, four visible
CP-mixing neutral Higgs bosons and a pair of oppositely charged Higgs
states. We found
significant differences between the possible mass spectra of the
CN2HDM type I and II. In type I the
overall spectra can range from heavy decoupled BSM particles with only
one SM-like scalar at $125\gev$ to very light spectra with the
heaviest of the four neutral scalars at $125\gev$. In type II, however,
the interplay of flavour and electroweak precision constraints always
leads to rather heavy mass spectra where the lightest neutral scalar
is always at $125\gev$.

Investigating the DM properties of the model, we found that in
either Yukawa type the model can easily account for 100\% of the observed relic
density. Due to a cancellation in the tree-level direct detection cross sections
the model at the same time remains completely out of reach of direct DM
searches. However, this does not make it impossible to probe the dark sector of
the CN2HDM. We found that for a Yukawa sector of type I the branching ratio of
the observed Higgs boson $h_{125}$ into a pair of DM particles can reach about
10.6\% after taking into account the LHC Higgs measurements. This is just below
the direct experimental upper limit for Higgs to invisible decays. With these
limits getting stronger and stronger, it becomes clear that in the future
higher-order electroweak corrections will have to be included in the prediction for the
Higgs-to-invisible branching ratio to be able to draw meaningful conclusions on
the still allowed Higgs coupling values. In type II the heavier mass spectra
also influence the allowed DM masses, such that DM masses below $m_{h_{125}}/2$
are extremely rare. The decay rates of the non-SM-like Higgs bosons into DM
particles can be quite large depending on their mass values. They reach 25 pb in
type I and 1.3 pb in type II which has an overall heavier mass spectrum. For
  many parameter points one or more of the non-SM-like Higgs bosons decay almost
  exclusively into invisible final states, which could be the only feasible decay
  channel to discover those particles.

The non-minimal Higgs sector of the model allows for a rich LHC phenomenology. In type I all four visible neutral Higgs
bosons can play the role of the discovered Higgs boson, meaning that \eg if the heaviest one, $H_4$, is the SM-like Higgs boson,
we have three visible light neutral Higgs bosons with masses below
$125\gev$ in the spectrum. We thus have a large variety of possible
Higgs decays and signatures. In particular, spectacular Higgs-to-Higgs
cascade decays with multiple Higgs bosons in the final state are
possible with rates above $1\fb$. In the type II version on the other
hand, the overall heavier Higgs spectra only allow for the lightest neutral non-DM Higgs boson to be the SM-like one. At the same time, since in the type II CN2HDM the up-
and down-type fermions couple to two different Higgs doublets the
SM-like Higgs boson can have larger CP-odd and singlet admixtures than
the corresponding Higgs in type I while still being in agreement with
all constraints. The CP-odd admixture can be up to 9\% and the singlet
admixture can reach 19\% compared with the
corresponding type I values of 7\% and 12\%, respectively. The combined
CP-odd-singlet admixture reaches 12.5\% in type I and 19\% in type
II\@. Both Yukawa types can predict deviations from the SM in the
signal strengths of $h_{125}$ that could be observed at the HL-LHC or
at a future collider. Distinguishing between the Yukawa types based on
signal strength measurements alone is challenging, but may be
possible, in particular if the signal rates are found to be larger than one.

With the CN2HDM we have presented an interesting benchmark model that can solve
some of the most pressing open questions in the SM. The next steps will be
further systematic investigations of its collider and DM phenomenology in order
to identify signatures to discover the whole Higgs spectrum, and signatures that
are specific for the model, \ie smoking gun signatures, to distinguish it from
other models. Observables that allow probing the amount of CP violation in the
model also warrant investigation. Finally, while the model offers all of the
required ingredients, it remains to be shown whether electroweak baryogenesis
can indeed be realised through a sufficiently strong first order electroweak
phase transition. The first step is done, the model is introduced and we have
provided and published the required tools so that anyone interested can study
the CN2HDM. Now further work is needed to corner the model experimentally and
theoretically.


\section*{Acknowledgements}
The authors would like to acknowledge Duarte Azevedo, Philipp Basler
and Rui Santos for fruitful discussions. The work of MM and SLW has
been supported by the Deutsche Forschungsgemeinschaft (DFG, German
Research Foundation) under grant 396021762 - TRR 257.
JM acknowledges support by the BMBF-Project 05H18VKCC1. The work of JW is funded by the Swedish Research Council, contract number 2016-0599.


\appendix
\section*{Appendix}

\section{The $4\times4$ Higgs Mixing Matrix} \label{Appendix:MixingMatrix}

\noindent
The orthogonal 4-dimensional mixing matrix is given by
($x,y=a,b,c,d$; $i,j=1,2,3,4$)
\beq
  R_{xy}= R_{ij} = R =
  \begin{pmatrix} R_{11} & R_{12} & R_{13} & R_{14} \\
                R_{21} & R_{22} & R_{23} & R_{24} \\
                R_{31} & R_{32} & R_{33} & R_{34} \\
                R_{41} & R_{42} & R_{32} & R_{44}
  \end{pmatrix},
\eeq
where $R$ is parametrised by the order and direction of
rotation of the angles $\alpha_1, \alpha_2, \alpha_3, \alpha_4,
  \alpha_5, \alpha_6$ about the $\{ x, y, z, w \}$ axes.

When we rotate in 4 dimensions, we rotate in each of the $x-y, x-z,
  x-w, y-z, y-w, z-w$ planes. We can modify our parametrisation of $R$
depending on which $\alpha_i$ characterises which plane of rotation
and in which order we choose to rotate about our axes. There is no
evident parametrisation in 4 dimensions, only that it is desirable to
obtain a parametrisation giving rise to 4 conditions that easily allow
us to describe the unique mixing matrix.

We choose the parametrisation
\begin{equation}
  R = \begin{aligned}[t]
     & R(\alpha_4, \{x, z\}) \cdot R(\alpha_5, \{x, y\}) \cdot R(-\alpha_3, \{x, w\})        \\
     & \cdot R(\alpha_6, \{y, z\}) \cdot R(-\alpha_2, \{y, w\}) \cdot R(\alpha_1, \{z, w\}),
  \end{aligned}
\end{equation} where $R(\alpha, \{p,q\})$ denotes a rotation in the plane $\{p,q\}$ characterised by $\alpha$. This choice yields the matrix comprised of the elements
\begin{subequations}
  \begin{align}
    R_{11} & = \caO \caT \caS                                                                                      \\
    R_{12} & = \caT \caS \saO                                                                                      \\
    R_{13} & = \caS \saT                                                                                           \\
    R_{14} & = -\saS                                                                                               \\
    R_{21} & = -\saO (\caTh \caF+\saTh \saF \saFi)+\caO (-\saT (\caF \saTh-\caTh \saF \saFi)-\caT \caFi \saF \saS) \\
    R_{22} & = \caO (\caTh \caF+\saTh \saF \saFi)+\saO (-\saT (\caF \saTh-\caTh \saF \saFi)-\caT \caFi \saF \saS)  \\
    R_{23} & = \caT (\caF \saTh-\caTh \saF \saFi)-\caFi \saT \saF \saS                                             \\
    R_{24} & = -\caFi \caS \saF                                                                                    \\
    R_{31} & = \caFi \saO \saTh+\caO (-\caTh \caFi \saT-\caT \saFi \saS)                                           \\
    R_{32} & = -\caO \caFi \saTh+\saO (-\caTh \caFi \saT-\caT \saFi \saS)                                          \\
    R_{33} & = \caT \caTh \caFi-\saT \saFi \saS                                                                    \\
    R_{34} & = -\caS \saFi                                                                                         \\
    R_{41} & = -\saO (\caTh \saF-\caF \saTh \saFi)+\caO (-\saT (\saTh \saF+\caTh \caF \saFi)+\caT \caF \caFi \saS) \\
    R_{42} & = \caO (\caTh \saF-\caF \saTh \saFi)+\saO (-\saT (\saTh \saF+\caTh \caF \saFi)+\caT \caF \caFi \saS)  \\
    R_{43} & = \caT (\saTh \saF+\caTh \caF \saFi)+\caF \caFi \saT \saS                                             \\
    R_{44} & = \caF \caFi \caS\,,
  \end{align}
\end{subequations}
where the $4\times4$ matrix reduces to the standard $3\times3$
rotation matrix as in Eq.~(2.15) of Ref.~\cite{Fontes:2017zfn} when
$\alpha_{4,5,6} \rightarrow 0$.
Without loss of generality, we choose to work in the convention where
\begin{align}
  -\pi/2 \leq \alpha_i \leq \pi/2\,,
\end{align}
which always yields $\cos{\alpha_i} \geq 0$.
The four conditions constraining the matrix are given by
\begin{align}
  R_{11}                                                       & = \caO \caT \caS > 0 \\
  R_{44}                                                       & = \caF \caFi \caS >0 \\
  R_{33} + \frac{R_{13} \cdot R_{14} \cdot R_{34}}{1-R_{14}^2} & = \caT \caT \caFi >0 \\
  \text{Det(R)}                                                & = 1\,.
\end{align}
These allow expressing the mixing angles in terms of the rotation matrix elements as
\begin{subequations}
  \begin{align}
    \alpha_6 & = \arcsin(-R_{14})\,                                                \\
    \alpha_2 & = \arcsin\left(\frac{R_{13}}{\caS} \right)                        \\
    \alpha_5 & = \arcsin \left( -\frac{R_{34}}{\caS} \right)                    \\
    \alpha_4 & = \arcsin \left( -\frac{R_{24}}{\caFi \caS }\right)              \\
    \alpha_1 & = \arcsin \left( \frac{R_{12}}{\caT \caS} \right)                \\
    \alpha_3 & = \arcsin \left( \frac{R_{31} \saO - R_{32} \caT}{\caFi} \right)\,,
  \end{align}
\end{subequations}
where each angle is written in terms of $\arcsin$, to avoid a
potential sign ambiguity of the angle that can arise from $\arccos$
functions (since $\cos(-x) = \cos(x)$).

\section{Relations between Lagrangian and Physical Parameters} \label{Appendix:LagrangianParameters}

Calculating the dependent Higgs mass values and reordering the mass set, we can then find the relations between the Lagrangian and the physical parameters of the model.
Defining 
\beq
X = R^T \mathrm{diag}(m_{H_1}^2,m_{H_2}^2,m_{H_3}^2 ,m_{H_4}^2 ) R
\eeq
for convenience, we find the quartic couplings,
\begin{align}
  \lambda_1 =     & \frac{X_{11} \cos\beta - \Re m_{12}^2 \sin\beta
                    }{v^2 \cos^3\beta}                \\
  \lambda_2 =     & \frac{X_{22}\sin\beta - \Re m_{12}^2
                    \cos\beta}{v^2 \sin^3\beta}                   \\
  \lambda_3 =     & \frac{2 m_{H^\pm}^2}{v^2} + \frac{X_{12} - \Re
                    m_{12}^2}{v^2 \sin\beta \cos\beta}  \\
  \lambda_4 =     & \frac{X_{44} - 2 m_{H^\pm}^2 }{v^2} + \frac{\Re
                    m_{12}^2}{v^2 \cos\beta \sin\beta}  \\
  \Re \lambda_5 = & \frac{\Re m_{12}^2}{v^2\sin\beta \cos\beta} -
                    \frac{X_{44}}{v^2}                    \\
  \Im \lambda_5 = & \frac{-2X_{24}}{v^2 \cos\beta}                                               \\
  \lambda_6 =     & \frac{4X_{33}}{v_s^2}                                                               \\
  \lambda_7 =     & \frac{2X_{13}}{v v_s \cos\beta}                                                    \\
  \lambda_8 =     & \frac{2X_{23}}{v v_s \sin\beta}\,.
\end{align}





\end{document}